\pdfoutput=1
\documentclass[12pt,a4paper]{article}
\usepackage{color}

\usepackage{ifthen} 
\newboolean{pdflatex}
\setboolean{pdflatex}{true} 

\newboolean{articletitles}
\setboolean{articletitles}{true} 

\newboolean{uprightparticles}
\setboolean{uprightparticles}{false} 

\def\paperauthors{LHCb collaboration} 
\def\paperasciititle{Measurement of the Lb -> Lambda(1520) mu mu differential branching fraction} 
\def\papertitle{Measurement of the \LbLambdastarmumu differential branching fraction} 
\def\paperkeywords{{High Energy Physics}, {LHCb}} 
\def\papercopyright{\the\year\ CERN for the benefit of the LHCb collaboration} 
\def\paperlicence{CC BY 4.0 licence}
\def\paperlicenceurl{https://creativecommons.org/licenses/by/4.0/}

\usepackage[top=1in, bottom=1.25in, left=1in, right=1in]{geometry}

\usepackage{bm}
\usepackage{float}
\usepackage{microtype}
\usepackage{lineno}  
\usepackage{xspace} 
\usepackage{caption} 

\usepackage{graphicx}  
\usepackage{color}
\usepackage{colortbl}
\graphicspath{{./figs/}} 

\usepackage{amsmath} 
\usepackage{amssymb}
\usepackage{amsfonts}
\usepackage{upgreek} 

\newcommand*\patchAmsMathEnvironmentForLineno[1]{%
\expandafter\let\csname old#1\expandafter\endcsname\csname #1\endcsname
\expandafter\let\csname oldend#1\expandafter\endcsname\csname
end#1\endcsname
 \renewenvironment{#1}%
   {\linenomath\csname old#1\endcsname}%
   {\csname oldend#1\endcsname\endlinenomath}%
}
\newcommand*\patchBothAmsMathEnvironmentsForLineno[1]{%
  \patchAmsMathEnvironmentForLineno{#1}%
  \patchAmsMathEnvironmentForLineno{#1*}%
}
\AtBeginDocument{%
\patchBothAmsMathEnvironmentsForLineno{equation}%
\patchBothAmsMathEnvironmentsForLineno{align}%
\patchBothAmsMathEnvironmentsForLineno{flalign}%
\patchBothAmsMathEnvironmentsForLineno{alignat}%
\patchBothAmsMathEnvironmentsForLineno{gather}%
\patchBothAmsMathEnvironmentsForLineno{multline}%
\patchBothAmsMathEnvironmentsForLineno{eqnarray}%
}

\usepackage{hyperxmp}

\usepackage[pdftex,
            pdfauthor={\paperauthors},
            pdftitle={\paperasciititle},
            pdfkeywords={\paperkeywords},
            pdfcopyright={Copyright (C) \papercopyright},
            pdflicenseurl={\paperlicenceurl}]{hyperref}

\usepackage[colorinlistoftodos,textsize=scriptsize]{todonotes}

\usepackage[bottom,flushmargin,hang,multiple]{footmisc}

\usepackage[all]{hypcap} 

\usepackage{xspace} 
\usepackage{upgreek}

\def\lhcb   {\mbox{LHCb}\xspace}

\def\MagUp {\mbox{\em Mag\kern -0.05em Up}\xspace}

\ifthenelse{\boolean{uprightparticles}}%
{

 \def\Pmu         {\ensuremath{\upmu}\xspace}                 
 \def\Pnu         {\ensuremath{\upnu}\xspace}                 
                  
 \def\Ppi         {\ensuremath{\uppi}\xspace}

 \def\Pphi        {\ensuremath{\upphi}\xspace}

 \def\Ppsi        {\ensuremath{\uppsi}\xspace}

 \def\PDelta      {\ensuremath{\Delta}\xspace}                 
 \def\PXi         {\ensuremath{\Xi}\xspace}                 
 \def\PLambda     {\ensuremath{\Lambda}\xspace}                 
 \def\PSigma      {\ensuremath{\Sigma}\xspace}                 
 \def\POmega      {\ensuremath{\Omega}\xspace}                 
 \def\PUpsilon    {\ensuremath{\Upsilon}\xspace}
 \let\oldPi\Pi
 \def\PPi         {\ensuremath{\oldPi}\xspace}

 \def\PB      {\ensuremath{\mathrm{B}}\xspace}                 
                  
 \def\PD      {\ensuremath{\mathrm{D}}\xspace}

 \def\PJ      {\ensuremath{\mathrm{J}}\xspace}                 
 \def\PK      {\ensuremath{\mathrm{K}}\xspace}

 \def\Pb      {\ensuremath{\mathrm{b}}\xspace}                 
 \def\Pc      {\ensuremath{\mathrm{c}}\xspace}

 \def\Pi      {\ensuremath{\mathrm{i}}\xspace}

 \def\Pp      {\ensuremath{\mathrm{p}}\xspace}

 \def\Ps      {\ensuremath{\mathrm{s}}\xspace}

 \def\thebaroffset{0.0em}
}
{

 \def\Pmu         {\ensuremath{\mu}\xspace}                 
 \def\Pnu         {\ensuremath{\nu}\xspace}                 
                  
 \def\Ppi         {\ensuremath{\pi}\xspace}

 \def\Pphi        {\ensuremath{\phi}\xspace}

 \def\Ppsi        {\ensuremath{\psi}\xspace}                 
                  
 \mathchardef\PDelta="7101
 \mathchardef\PXi="7104
 \mathchardef\PLambda="7103
 \mathchardef\PSigma="7106
 \mathchardef\POmega="710A
 \mathchardef\PUpsilon="7107
 \mathchardef\PPi="7105
                  
 \def\PB      {\ensuremath{B}\xspace}                 
                  
 \def\PD      {\ensuremath{D}\xspace}

 \def\PJ      {\ensuremath{J}\xspace}                 
 \def\PK      {\ensuremath{K}\xspace}

 \def\Pb      {\ensuremath{b}\xspace}                 
 \def\Pc      {\ensuremath{c}\xspace}

 \def\Pi      {\ensuremath{i}\xspace}

 \def\Pp      {\ensuremath{p}\xspace}

 \def\Ps      {\ensuremath{s}\xspace}

 \def\thebaroffset{0.18em}
}
\newcommand{\offsetoverline}[2][\thebaroffset]{\kern #1\overline{\kern -#1 #2}}%

\makeatletter
\ifcase \@ptsize \relax
  \newcommand{\miniscule}{\@setfontsize\miniscule{4}{5}}
\or
  \newcommand{\miniscule}{\@setfontsize\miniscule{5}{6}}
\or
  \newcommand{\miniscule}{\@setfontsize\miniscule{5}{6}}
\fi
\makeatother

\DeclareRobustCommand{\optbar}[1]{\shortstack{{\miniscule (\rule[.5ex]{1.25em}{.18mm})}
  \\ [-.7ex] $#1$}}

\def\muon       {{\ensuremath{\Pmu}}\xspace}
\def\mup        {{\ensuremath{\Pmu^+}}\xspace}
\def\mun        {{\ensuremath{\Pmu^-}}\xspace} 

\def\mumu       {{\ensuremath{\Pmu^+\Pmu^-}}\xspace}

\def\ellell     {\ensuremath{\ell^+ \ell^-}\xspace}

\def\neu        {{\ensuremath{\Pnu}}\xspace}
\def\neub       {{\ensuremath{\overline{\Pnu}}}\xspace}

\def\squark    {{\ensuremath{\Ps}}\xspace}

\def\cquark    {{\ensuremath{\Pc}}\xspace}

\def\bquark    {{\ensuremath{\Pb}}\xspace}

\def\pion   {{\ensuremath{\Ppi}}\xspace}

\def\pip    {{\ensuremath{\pion^+}}\xspace}
\def\pim    {{\ensuremath{\pion^-}}\xspace}

\def\kaon    {{\ensuremath{\PK}}\xspace}

\def\KorKbar {\kern \thebaroffset\optbar{\kern -\thebaroffset \PK}{}\xspace}

\def\Kp      {{\ensuremath{\kaon^+}}\xspace}
\def\Km      {{\ensuremath{\kaon^-}}\xspace}

\def\KS      {{\ensuremath{\kaon^0_{\mathrm{S}}}}\xspace}

\def\Kstarz  {{\ensuremath{\kaon^{*0}}}\xspace}

\def\Kstar   {{\ensuremath{\kaon^*}}\xspace}

\def\D       {{\ensuremath{\PD}}\xspace}

\def\DorDbar {\kern \thebaroffset\optbar{\kern -\thebaroffset \PD}\xspace}
\def\Dz      {{\ensuremath{\D^0}}\xspace}

\def\Dp      {{\ensuremath{\D^+}}\xspace}
\def\Dm      {{\ensuremath{\D^-}}\xspace}

\def\DpDm    {\ensuremath{\Dp {\kern -0.16em \Dm}}\xspace}

\def\B       {{\ensuremath{\PB}}\xspace}

\def\BorBbar {\kern \thebaroffset\optbar{\kern -\thebaroffset \PB}\xspace}
\def\Bz      {{\ensuremath{\B^0}}\xspace}

\def\Bd      {{\ensuremath{\B^0}}\xspace}

\def\BdorBdbar {\kern \thebaroffset\optbar{\kern -\thebaroffset \Bd}\xspace}
\def\Bu      {{\ensuremath{\B^+}}\xspace}
\def\Bub     {{\ensuremath{\B^-}}\xspace}
\def\Bp      {{\ensuremath{\Bu}}\xspace}
\def\Bm      {{\ensuremath{\Bub}}\xspace}

\def\Bs      {{\ensuremath{\B^0_\squark}}\xspace}

\def\BsorBsbar {\kern \thebaroffset\optbar{\kern -\thebaroffset \Bs}\xspace}

\def\jpsi     {{\ensuremath{{\PJ\mskip -3mu/\mskip -2mu\Ppsi}}}\xspace}
\def\psitwos  {{\ensuremath{\Ppsi{(2S)}}}\xspace}

\def\Y#1S{\ensuremath{\PUpsilon{(#1S)}}\xspace}

\def\proton      {{\ensuremath{\Pp}}\xspace}

\def\Lz          {{\ensuremath{\PLambda}}\xspace}

\def\LorLbar     {\kern \thebaroffset\optbar{\kern -\thebaroffset \PLambda}\xspace}

\def\Lc          {{\ensuremath{\Lz^+_\cquark}}\xspace}

\def\Lb           {{\ensuremath{\Lz^0_\bquark}}\xspace}

\newcommand{\decay}[2]{\ensuremath{#1\!\to #2}\xspace} 

\def\to                 {\ensuremath{\rightarrow}\xspace}

\def\qsq       {{\ensuremath{q^2}}\xspace}

\def\CP                {{\ensuremath{C\!P}}\xspace}

\def\AT#1     {\ensuremath{A_{\mathrm{T}}^{#1}}\xspace}           

\def\C#1      {\ensuremath{\mathcal{C}_{#1}}\xspace}                       
\def\Cp#1     {\ensuremath{\mathcal{C}_{#1}^{'}}\xspace}                    
\def\Ceff#1   {\ensuremath{\mathcal{C}_{#1}^{\mathrm{(eff)}}}\xspace}        
\def\Cpeff#1  {\ensuremath{\mathcal{C}_{#1}^{'\mathrm{(eff)}}}\xspace}       
\def\Ope#1    {\ensuremath{\mathcal{O}_{#1}}\xspace}                       
\def\Opep#1   {\ensuremath{\mathcal{O}_{#1}^{'}}\xspace}                    

\newcommand{\aunit}[1]{\ensuremath{\text{\,#1}}}       

\newcommand{\tev}{\aunit{Te\kern -0.1em V}\xspace}
\newcommand{\gev}{\aunit{Ge\kern -0.1em V}\xspace}
\newcommand{\mev}{\aunit{Me\kern -0.1em V}\xspace}
\newcommand{\kev}{\aunit{ke\kern -0.1em V}\xspace}
\newcommand{\ev}{\aunit{e\kern -0.1em V}\xspace}
 
\newcommand{\mevc}{\ensuremath{\aunit{Me\kern -0.1em V\!/}c}\xspace}
\newcommand{\gevc}{\ensuremath{\aunit{Ge\kern -0.1em V\!/}c}\xspace}
\newcommand{\mevcc}{\ensuremath{\aunit{Me\kern -0.1em V\!/}c^2}\xspace}
\newcommand{\gevcc}{\ensuremath{\aunit{Ge\kern -0.1em V\!/}c^2}\xspace}
\newcommand{\gevgevcccc}{\ensuremath{\gev^2\!/c^4}\xspace} 

\def\fb   {\ensuremath{\aunit{fb}}\xspace}
\def\invfb   {\ensuremath{\fb^{-1}}\xspace}

\def\gsim{{~\raise.15em\hbox{$>$}\kern-.85em
          \lower.35em\hbox{$\sim$}~}\xspace}
\def\lsim{{~\raise.15em\hbox{$<$}\kern-.85em
          \lower.35em\hbox{$\sim$}~}\xspace}

\def\sPlot{\mbox{\em sPlot}\xspace}

\def\pt         {\ensuremath{p_{\mathrm{T}}}\xspace}

\def\evtgen     {\mbox{\textsc{EvtGen}}\xspace}

\def\geant      {\mbox{\textsc{Geant4}}\xspace}

\def\photos     {\mbox{\textsc{Photos}}\xspace}

\def\pythia     {\mbox{\textsc{Pythia}}\xspace}

\def\tell1  {TELL1\xspace}
\def\ukl1   {UKL1\xspace}

\def\LbLambdamumu{{\ensuremath{\Lb\to \PLambda \mumu}}\xspace}

\def\Lbpkmumu{{\ensuremath{\Lb\to \proton\Km \mumu}}\xspace}

\def\LbLambdastarmumu{{\ensuremath{\Lb\to \PLambda(1520) \mumu}}\xspace}
\def\LbLambdastarpkmumu{{\ensuremath{\Lb\to \PLambda(1520)(\to\proton\Km) \mumu}}\xspace}
\def\Lambdastarmumu{{\ensuremath{ \PLambda(1520) \mumu}}\xspace}
\def\pkjpsi{{\ensuremath{\proton\Km J/\Ppsi}}\xspace}

\def\Lbpkmumu{{\ensuremath{\Lb\to \proton\Km \mumu}}\xspace}

\def\Lbpkjpsi{{\ensuremath{\Lb\to \proton\Km J/\Ppsi}}\xspace}

\def\Bskkjpsi  {\decay{\Bs}{\Kp \Km \jpsi}}

\def\Bdkpijpsi{\decay{\Bd}{\Kp\pim \jpsi}}

\def\btosll{{\ensuremath{\bquark\to \squark  \ellell}}\xspace}

\def\Lbpkll{{\ensuremath{\Lb\to p K^- \ellell}}\xspace}

\def\lstar{{\ensuremath{\PLambda}}\xspace}
\def\lsig{{\ensuremath{\PLambda(1520)}}\xspace}
\def\llow{{\ensuremath{\PLambda(1405)}}\xspace}

\def\lmid{{\ensuremath{\PLambda(1600)}}\xspace}
\def\lhigh{{\ensuremath{\PLambda(1800)}}\xspace}
\def\JpsiDecay{{\ensuremath{\jpsi\to \mumu}}\xspace}
\def\lsigdecay{{\ensuremath{\PLambda(1520)\to \proton\Km}}\xspace}
\def\pkmumu{{\ensuremath{\proton\Km \mumu}}\xspace}

\def\qsqmin      {{\ensuremath{q^{2}_{\rm min}}}\xspace}
\def\qsqmax      {{\ensuremath{q^{2}_{\rm max}}}\xspace}

\newcommand{\lhcborcid}[1]{\href{https://orcid.org/#1}{\hspace*{0.1em}\raisebox{-0.45ex}{\includegraphics[width=1em]{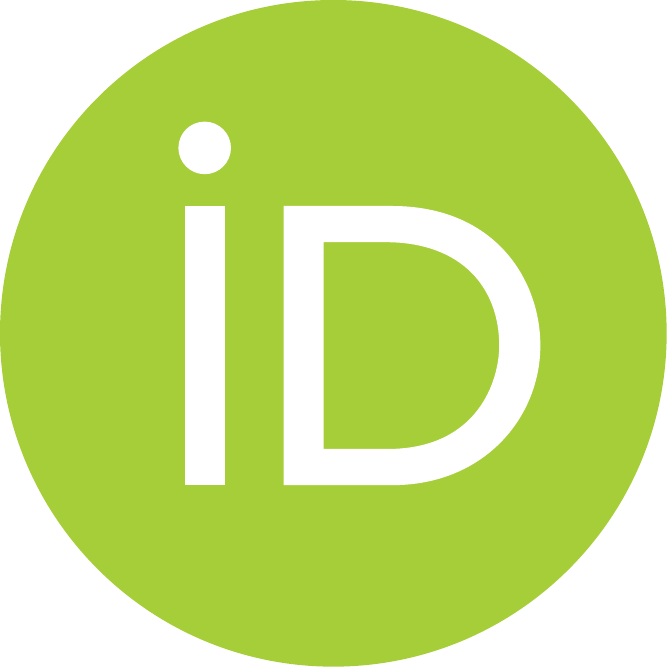}}}}

\def\sPlot{\mbox{\em sPlot}\xspace}

\usepackage{multirow} 
\usepackage{booktabs} 
\usepackage{rotating}

\usepackage{cite} 
\usepackage{mciteplus}

\usepackage{longtable} 

\begin{document}

\renewcommand{\thefootnote}{\fnsymbol{footnote}}
\setcounter{footnote}{1}


\begin{titlepage}
\pagenumbering{roman}

\vspace*{-1.5cm}
\centerline{\large EUROPEAN ORGANIZATION FOR NUCLEAR RESEARCH (CERN)}
\vspace*{1.5cm}
\noindent
\begin{tabular*}{\linewidth}{lc@{\extracolsep{\fill}}r@{\extracolsep{0pt}}}
\ifthenelse{\boolean{pdflatex}}
{\vspace*{-1.5cm}\mbox{\!\!\!\includegraphics[width=.14\textwidth]{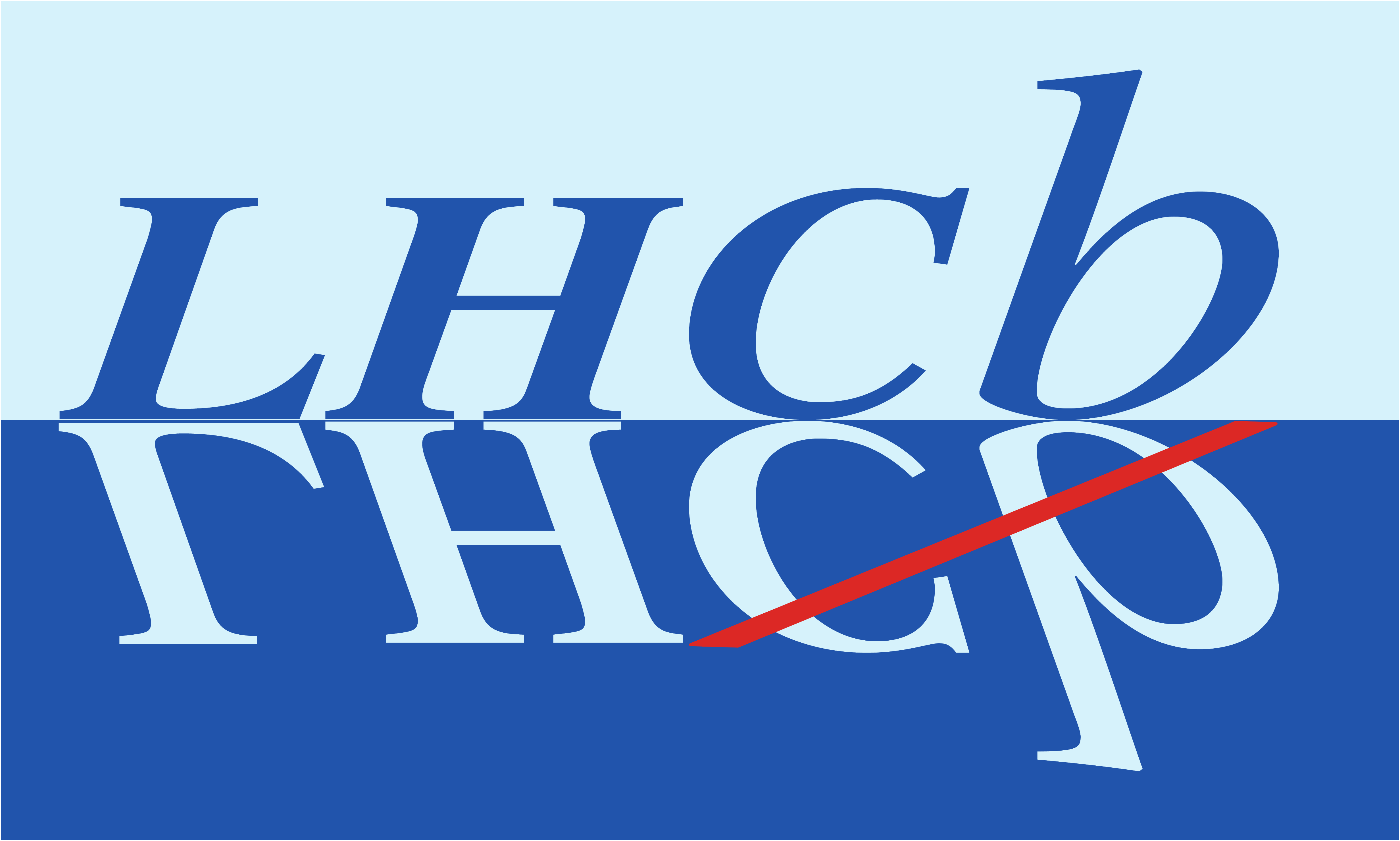}} & &}%
{\vspace*{-1.2cm}\mbox{\!\!\!\includegraphics[width=.12\textwidth]{figs/lhcb-logo.eps}} & &}%
\\
 & & CERN-EP-2023-007 \\  
 & & LHCb-PAPER-2022-050 \\  
 & & 23 February 2023 \\ 
 & & \\
\end{tabular*}

\vspace*{4.0cm}

{\normalfont\bfseries\boldmath\huge
\begin{center}
  \papertitle 
\end{center}
}

\vspace*{2.0cm}


\begin{center}
\paperauthors\footnote{Authors are listed at the end of this paper.} 
\end{center}

\vspace{\fill}

\begin{abstract}
  \noindent
  The branching fraction of the rare decay $\LbLambdastarmumu$ is measured for the first time, in the squared dimuon mass intervals $q^2$, excluding the \jpsi and \psitwos regions. The data sample analyzed was collected by the LHCb experiment 
  at center-of-mass energies of 7, 8, and 13\tev, corresponding to a total integrated  luminosity of 9\invfb. 
  The result in the highest $\qsq$ interval, $\qsq >$15.0\gevgevcccc, where theoretical predictions have the smallest model dependence, agrees with the predictions.

\end{abstract}

\vspace*{2.0cm}

\begin{center}
Published in Phys. Rev. Lett. 131 (2023) 151801

\end{center}

\vspace{\fill}

{\footnotesize 
\centerline{\copyright~\papercopyright. \href{\paperlicenceurl}{\paperlicence}.}}
\vspace*{2mm}

\end{titlepage}


\newpage
\setcounter{page}{2}
\mbox{~}


\renewcommand{\thefootnote}{\arabic{footnote}}
\setcounter{footnote}{0}

\cleardoublepage


\pagestyle{plain} 
\setcounter{page}{1}
\pagenumbering{arabic}


\noindent The standard model (SM) of particle physics provides at present the best description of fundamental particles and their interactions. However, it is unable to explain the dominance of matter over antimatter or the patterns of the interaction strengths of the elementary particles. Physics beyond the SM (BSM) is needed to address these limitations.

One way of searching for BSM physics is to study the flavor changing neutral-current transition \btosll, which proceeds through electroweak loop diagrams in the SM, while a sizeable contribution could be introduced by BSM physics~\cite{Das:2018sms,Langacker:2000ju,Biswas:2022fyb}. 
Such decays have been studied in the \B-meson sector by measuring branching fractions~\cite{LHCb-PAPER-2013-043,LHCb-PAPER-2014-006,LHCb-PAPER-2016-012,LHCb-PAPER-2021-014} and angular distributions~\cite{CDF:2011tds, CMS:2015bcy, CMS:2017rzx,CMS:2018qih,ATLAS:2018gqc,CMS:2020oqb, LHCb-PAPER-2015-023,LHCb-PAPER-2015-051,Belle:2016fev,LHCb-PAPER-2020-041,LHCb-PAPER-2021-022} and testing lepton flavor universality ~\cite{BaBar:2012mrf, LHCb-PAPER-2019-009, BELLE:2019xld, Belle:2019oag,LHCb-PAPER-2021-038,LHCb:2022qnv,LHCb:2022zom}.
Similar to \B-meson decays, the study of \B-baryon decays offers a multitude of observables that are analogous to those typically measured in \B-meson decays, including charge-parity (\CP) asymmetries. 
Owing to the half-integer spin, the \B-baryon decays offer an even richer angular structure than the \B-meson decays~\cite{Feldmann:2021zta}. 

The differential branching fraction and angular observables of the \LbLambdamumu decay were analyzed by the LHCb Collaboration~\cite{LHCb-PAPER-2015-009}, and the measured values can be described by recent theoretical calculations~\cite{Detmold:2016pkz,Blake:2019guk}.
The \Lbpkmumu decay was first observed by the \lhcb Collaboration~\cite{LHCb-PAPER-2016-059}, and a search for \CP violation was performed. A test of lepton flavor universality in the 
decay \Lbpkll was carried out by the LHCb Collaboration~\cite{LHCb-PAPER-2019-040}, 
and the result was found to be consistent with SM predictions. In the aforementioned measurements, there are various contributions of excited $\PLambda$ baryon resonances to the $p\Km$ final state, among which the \lsig stands out as having a relatively narrow width of 16\mev~\cite{Sarantsev:2019xxm}. In contrast to the ground state $\PLambda$, which has a spin parity of $J^P=(1/2)^-$, the excited \lsig state has a spin parity of $J^P=(3/2)^-$, providing complementary information on potential new physics effects in the \btosll transitions~\cite{Meinel:2020owd}.

This Letter reports the first measurement of the differential branching fraction of the \LbLambdastarmumu decay in intervals of the squared dimuon mass, \qsq, with the \lsig baryon reconstructed through its \lsigdecay decay. The inclusion of charge-conjugate processes is implied throughout this Letter.
The more abundant tree-level decay \Lbpkjpsi, with a well-measured branching fraction~\cite{PDG2022}, is used for normalization.
The measurements are performed using proton-proton ($pp$) collision data corresponding to an integrated luminosity of 9\invfb recorded by the \lhcb experiment at center-of-mass energies of 7, 8 and 13\tev.

The \lhcb detector~\cite{LHCb-DP-2008-001,LHCb-DP-2014-002} is a single-arm forward spectrometer covering the \mbox{pseudorapidity} range $2<\eta <5$, designed for the study of particles containing \bquark or \cquark quarks. 
The online event selection is performed by a trigger~\cite{LHCb-DP-2012-004,LHCb-DP-2019-001}, which consists of a hardware stage, based on information from the calorimeters and muon systems~\cite{LHCb-DP-2013-001}, followed by a software stage, which applies a full event reconstruction.
Simulated events are used to develop the candidate selection and to estimate the corresponding efficiency for the signal and normalization modes.
In the simulation, $pp$ collisions are generated using
\pythia8~\cite{Sjostrand:2007gs,Sjostrand:2006za} 
with a specific \lhcb configuration~\cite{LHCb-PROC-2010-056}. Decays of unstable particles are described by \evtgen~\cite{Lange:2001uf}, in which final-state radiation is generated using \photos~\cite{Golonka:2005pn}. The interaction of the generated particles with the detector, and its response, are simulated using the \geant toolkit~\cite{Allison:2006ve, *Agostinelli:2002hh} as described in Ref.~\cite{LHCb-PROC-2011-006}. The 
\LbLambdastarmumu and \Lbpkjpsi decays are simulated following a uniform phase-space model, and the  intermediate resonant structures in the \Lbpkjpsi decay
are corrected based on the amplitude analysis in Ref.~\cite{LHCb-PAPER-2015-029}. The \Lb lifetime in the simulation is corrected to its known value~\cite{PDG2022}.  In addition, the detector occupancy and \Lb transverse momentum, $\pt(\Lb)$, distributions of all simulated samples involving \Lb decays are corrected for discrepancies between the simulation and data, using \Lbpkjpsi samples.

Candidate \lsig baryon decays are reconstructed from two oppositely charged tracks identified as a proton and a kaon originating from a common vertex. No requirement on the $m(p\Km)$ mass is applied. 
The \LbLambdastarpkmumu decay is reconstructed by combining a \lsig candidate with two oppositely charged tracks identified as muons. Dimuon pairs having mass squared \qsq values around the \jpsi ($8.0<\qsq<11.0\gevgevcccc$) and \psitwos ($12.5<\qsq<15.0\gevgevcccc$) resonances are vetoed from the signal decay sample, while the candidates from the \Lbpkjpsi decay are used for normalization. The background is further suppressed by requirements on the quality of the \Lb decay vertex, the flight distance significance of the \Lb candidate, the compatibility of the \Lb candidate to come from the primary $\proton\proton$ interaction vertex (PV), and the separation of the final-state tracks from the PV. For the \Lbpkjpsi decay, the mass of the \Lb candidate is recalculated with the \jpsi meson mass constrained to its known value~\cite{PDG2022}, leading to an improvement on the \Lb mass resolution.

Various vetoes on hadron and muon masses reject peaking backgrounds originating from misidentified \bquark-hadron decays, by recalculating the mass of the four-track combination under alternative particle hypotheses and removing candidates in the relevant mass range. 
Background candidates in the \LbLambdastarpkmumu sample can originate from a few different sources. These include the $\Bs\to\phi(1020)(\to \Kp\Km)\mumu$ decay, where a kaon is misidentified as a proton; contamination from $\Bm\to\Km\mumu$ decays combined with a random additional proton; and the $\Lb\to\proton\Dz(\to\Km\pip)\pim$ decay, where both pions are misidentified as muons.
The same vetoes on the $\Bs$ and $\Bm$ decays are used to reject backgrounds in the \Lbpkjpsi data sample. Owing to the narrow \jpsi mass window requirement, the $\Lb\to\proton\Dz(\to\Km\pip)\pim$ 
background is negligible for the \Lbpkjpsi decay, and therefore the veto is not applied.
The \jpsi contamination in the signal decay sample with a muon 
misidentified as a hadron is found to be negligible. These vetoes retain about 91\% of the signal while strongly suppressing all these background sources. The background from $\Lb\to\Lc(\to\proton\Km X)Y$ decays, where $X$ and $Y$ can represent either a $\mun\neub_{\muon}\, (\mup\neu_{\muon})$ pair or a pion, are verified to be negligible in the \LbLambdastarpkmumu signal sample.

In order to increase the signal purity, a multivariate classification is employed using a boosted decision tree (BDT)~\cite{Breiman,Roe:2004na} algorithm implemented in the TMVA package~\cite{Hocker:2007ht}. To train this classifier, simulated \LbLambdastarpkmumu candidates are used as the signal proxy, and candidates lying in the upper \proton\Km\mumu mass sideband (6.0--6.8\gevcc) adjacent to the signal region are used as background proxy. The variables used in the BDT training include kinematical and topological properties of the final state or intermediate particles. 
The requirement on the multivariate output 
is optimized by maximizing the magnitude of $N_{S}/\sqrt{N_{S}+N_{B}}$, where $N_{S}$ and $N_{B}$ are the expected number of \LbLambdastarmumu signal and background candidates underneath the signal peak, respectively. 
The same BDT selection is applied to the \Lbpkjpsi normalization mode.

With the full selection applied, the differential branching fraction of the \LbLambdastarmumu decay is determined in intervals of \qsq, relative to the \Lbpkjpsi normalization mode, according to
\begin{align}\label{eq:dbf}
\begin{split}
 \left\{\frac{d\mathcal{B}[\LbLambdastarmumu]}{d\qsq}\right\}^{\qsqmax}_{\qsqmin}= \frac{1}{(\qsqmax - \qsqmin)} & \frac{\mathcal{B}(\Lbpkjpsi)\mathcal{B}(\JpsiDecay)}{\mathcal{B}[\lsigdecay]} \\
 & \times   \frac{N_{\Lambdastarmumu}}{N_{\pkjpsi}}\frac{\varepsilon_{\pkjpsi}}{\varepsilon_{\Lambdastarmumu}},
\end{split}
\end{align}
where $N_{\pkjpsi}$ and $\varepsilon_{\pkjpsi}$ are the yields and efficiencies of the normalization mode, and $N_{\Lambdastarmumu}$ and $\varepsilon_{\Lambdastarmumu}$ indicate the corresponding parameters for the signal mode in the $\qsqmin$--$\qsqmax$ interval.
The branching fractions $\mathcal{B}(\Lbpkjpsi)$, $\mathcal{B}(\JpsiDecay)$ and $\mathcal{B}[\lsigdecay]$ are $(3.2\pm 0.6)\times 10^{-4}$, (5.961$\pm$0.033)\% and ($22.5\pm 0.5$)\%, respectively~\cite{PDG2022}. 

To determine the geometrical acceptance and the efficiencies of the trigger, reconstruction, and particle identification (PID) requirements, we utilize the simulation and apply corrections based on unbiased control samples selected from real data. These corrections are applied to simulated samples of the signal and normalization modes to refine their accuracy.
The PID efficiencies for each channel are calculated from calibration data samples of muons, proton, kaons and pions~\cite{LHCb-DP-2018-001}, and are applied as per-candidate weights to the simulation. 
Similarly, the trigger efficiency is corrected by comparing the efficiency in data and simulation as a function of the \pt of the muons in the normalization mode.  Finally, the relative efficiencies for different data-taking periods are combined according to the integrated luminosity times the production rate of the \Lb baryon to account for the variation of data-taking conditions.

\begin{figure}
  \centering
  \includegraphics[width=0.49\linewidth]{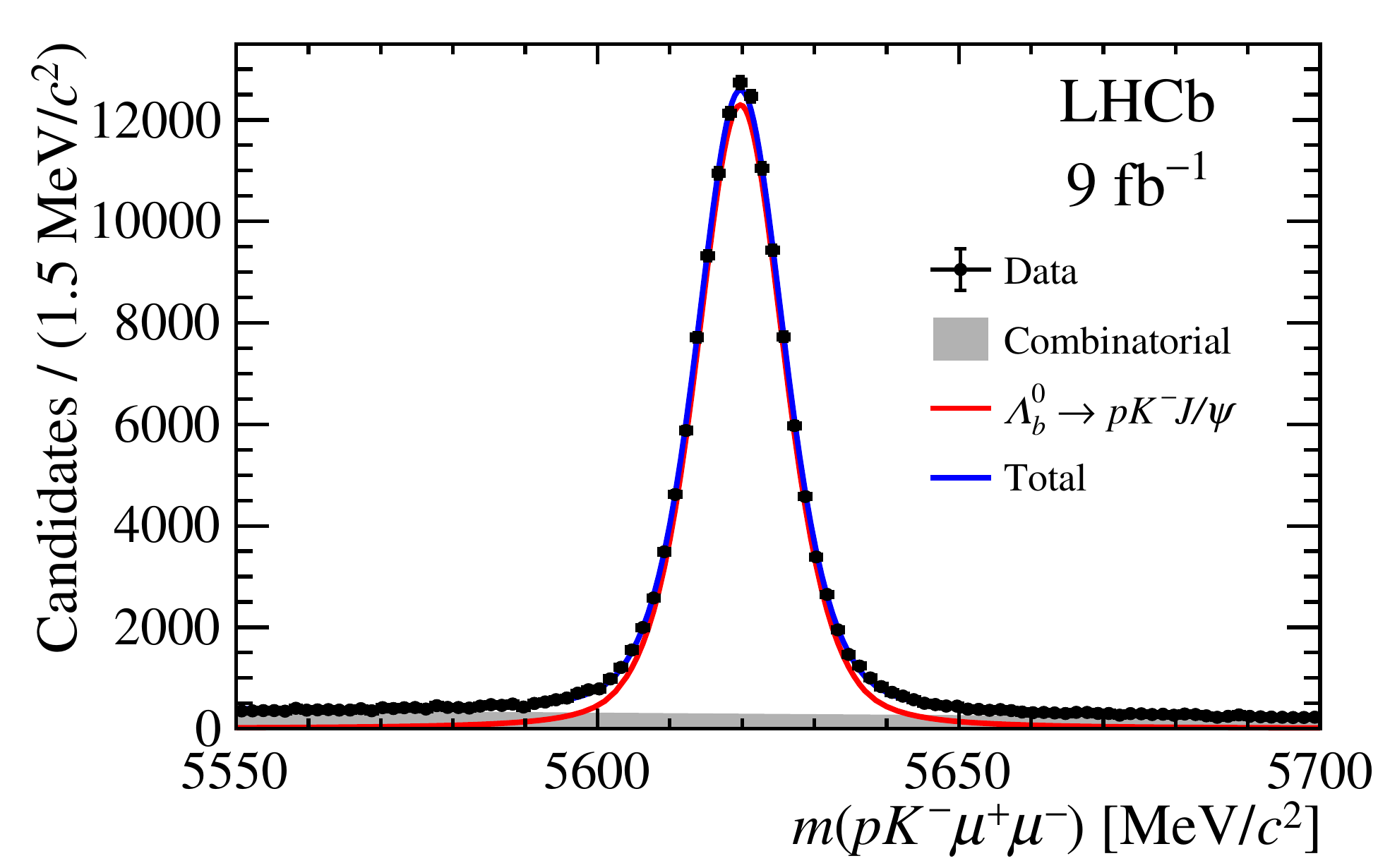}
  \includegraphics[width=0.49\linewidth]{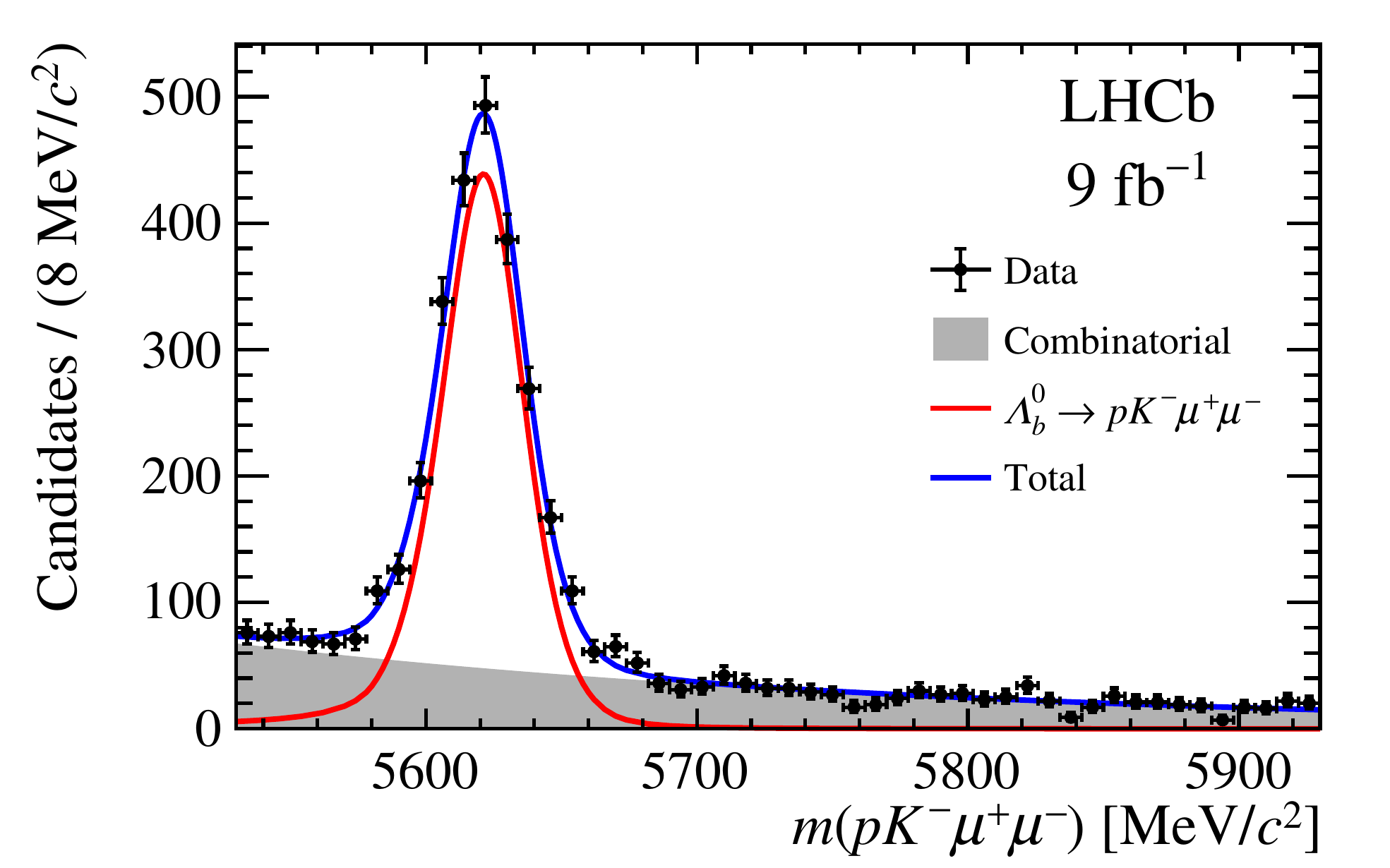}
  \caption{Mass distribution for the (left) \Lbpkjpsi and (right) \Lbpkmumu, integrated over the considered \qsq intervals. Fit results are overlaid.}
  \label{fig:lbfit}
\end{figure}

The yield of the $\Lbpkjpsi$ normalization mode is determined using an extended unbinned maximum-likelihood fit to the $\pkjpsi$ mass distribution with the \jpsi meson mass constrained to its known value~\cite{PDG2022}. The signal is modeled by a Hypatia function~\cite{Santos:2013gra}. The tail parameters are determined from simulation, while the peak position and resolution are allowed to vary freely in the fit to data.  The combinatorial background is modeled using an exponential function with the slope allowed to vary freely. The $m(\pkmumu)$  distribution of the selected $\Lbpkjpsi$ candidates is shown in Fig.~\ref{fig:lbfit} (left), with the fit results overlaid. 
The \Lbpkjpsi yield is found to be $N_{\pkjpsi} = 137\,900 \pm 405$, where the uncertainty is statistical only.

For the $\LbLambdastarmumu$ decay, we perform a simultaneous extended unbinned maximum-likelihood fit to the \pkmumu mass distribution in different intervals of \qsq. The mean of the signal peak is shared among the different samples. 
The same signal and background models as for the $\Lbpkjpsi$ normalization mode are used, while the tail parameters of the signal model are determined from the simulated $\LbLambdastarmumu$ sample. The $m(\pkmumu)$ resolution parameter for the signal component is fixed to the value obtained from a fit to the normalization mode. 
Figure~\ref{fig:lbfit} (right) shows the $m(\pkmumu)$ distribution of the full data sample, integrated over all the considered \qsq intervals. The \Lbpkmumu signal yield is found to be $N_{\Lbpkmumu}= 2\,250 \pm 57$, where the uncertainty is statistical only.
The corresponding figures for all the $\qsq$ ranges are available as Supplemental Material~\cite{sup}. 

\begin{figure}
  \centering
  \includegraphics[width=0.49\linewidth]{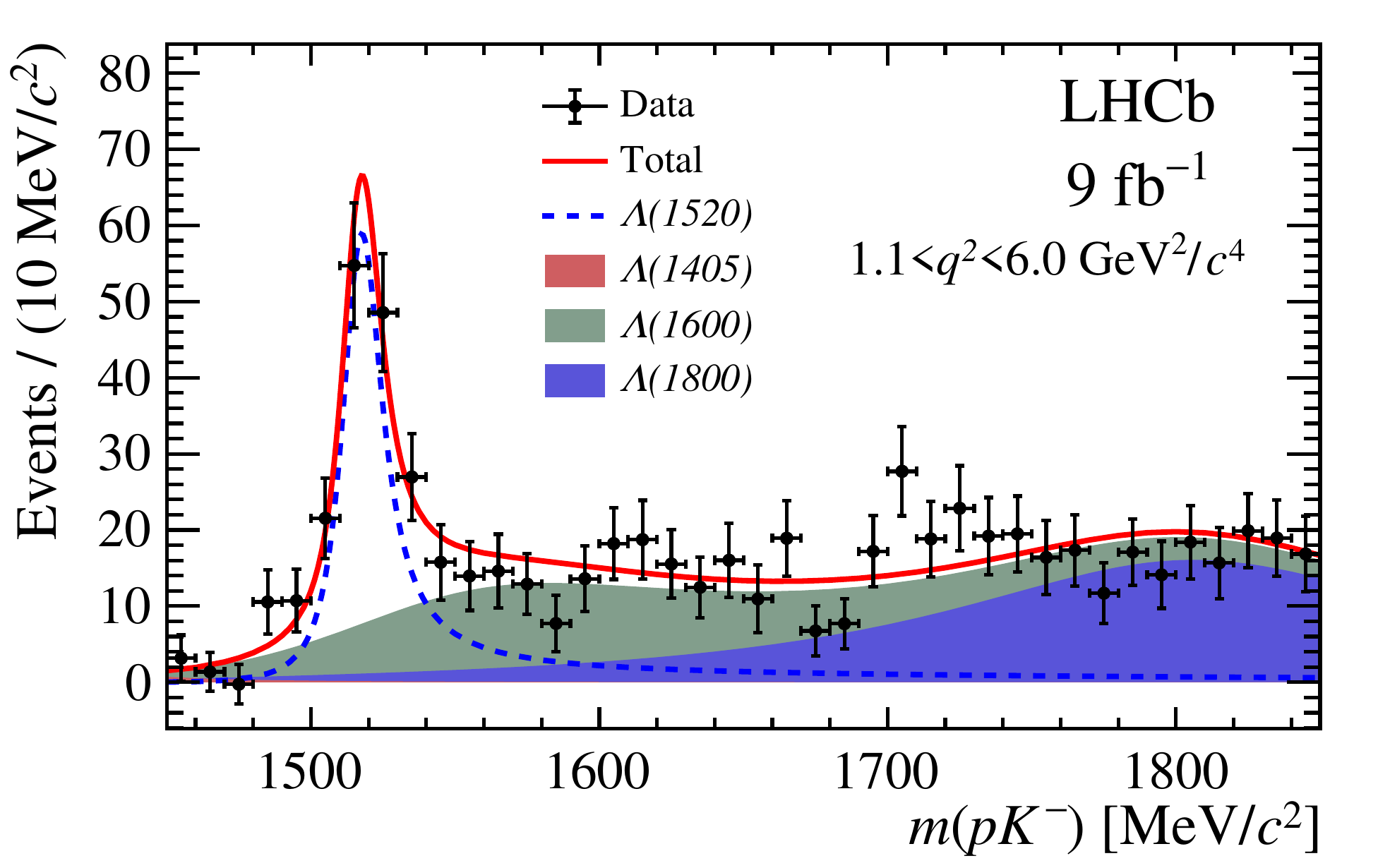}
  \includegraphics[width=0.49\linewidth]{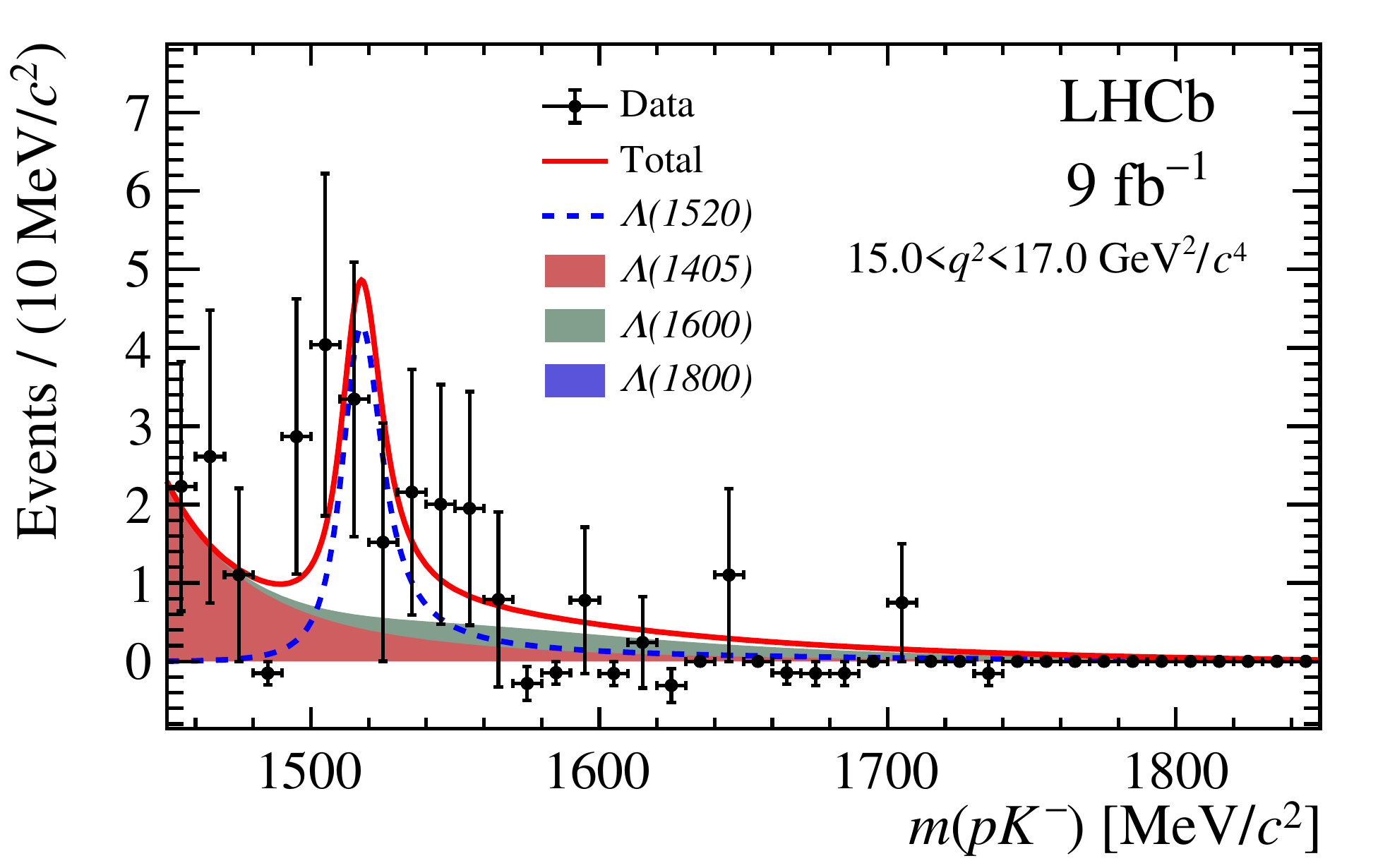}
  \caption{Distribution of the $\proton\Km$ mass, after background subtraction, for \LbLambdastarmumu signal candidates in the (left) $1.1<\qsq<6.0\gevgevcccc$ and (right) $15.0<\qsq<17.0\gevgevcccc$ regions. Fit results are overlaid.}
  \label{fig:xmfit}
\end{figure}

The yields of the \LbLambdastarmumu signal in the different \qsq bins are determined by maximizing the extended weighted log likelihood for unbinned $m(\proton\Km)$ distributions. The non-\Lb background is subtracted using the \sPlot technique, which utilizes the $m(\pkmumu)$ distribution as a discriminating variable~\cite{Pivk:2004ty,Xie:2009rka}.
The fit procedure is validated with pseudoexperiments, and the uncertainties of the \LbLambdastarmumu signal yields are corrected using the bootstrap method~\cite{Langenbruch:2019nwe}. A cross-check is performed using unbinned maximum-likelihood fits to the two-dimensional $m(\pkmumu)$
and $m(\proton\Km)$ distributions, which give consistent results. 
Considering the mass and width of all the $\lstar$ states~\cite{PDG2022} and their contributions to the background-subtracted $m(p\Km)$ spectra, 
the \llow, \lsig, \lmid, and \lhigh states are included in the nominal fits. 
The line shapes of these resonances are parametrized using relativistic Breit–Wigner functions. 
The efficiency as a function of $m(pK^{-})$ is determined from simulation and included in the fit model. 
As the \lsig resonance has a width that is comparable to the experimental resolution, the line shape is convolved with a Gaussian resolution function. 
The width of the Gaussian is taken from the simulation. 
In the fits to the background-subtracted $m(pK^-)$ distributions, the width and mass of all the \lstar resonances are fixed according to the world's best results~\cite{Hassanvand:2012dn,Sarantsev:2019xxm}. 
The $\PLambda(1670), \PLambda(1690), \PLambda(1820)$, and $\PLambda(1830)$ states, and interference effects are not included in the fits as these are found to be small, and a systematic uncertainty is included in that of the signal fit mode.
The background-subtracted $m(pK^{-})$ distribution in the \qsq regions 1.1--6.0$\gevgevcccc$ and 15.0--17.0$\gevgevcccc$ are shown in Fig.~\ref{fig:xmfit}.
The signal yields in all the $\qsq$ intervals are given in Table~\ref{tab:results}.

\begin{table}
    \centering
    \caption{Relative systematic uncertainties [in \%] of the differential branching fraction ratio measurement. The total uncertainty is obtained as the quadratic sum of the individual contributions.}
    \resizebox{\linewidth}{!}{
    \renewcommand*{\arraystretch}{1.2}
		\begin{tabular}{c cccccc}
			\toprule
   \noalign{\smallskip}
   \multirow{2}{*}{Source} & \multicolumn{6}{c}{\qsq interval $[\!\gevgevcccc]$} \\
			\cline{2-7}    & 0.1--3.0 & 3.0--6.0 & 6.0--8.0 & 11.0--12.5 & 15.0--17.0 & 1.1--6.0 \\
   \noalign{\smallskip}
			\midrule
								Signal fit model
		& 9.6 & 6.5 & 9.3 & 9.3 & 15.3 & 7.2    
			\\
				Normalization fit model                                         & 1.3     & 1.3     & 1.3     & 1.3  & 1.3 & 1.3
			\\

						Hardware trigger 
			& 0.3     & 0.5     & 0.2      & 0.1   & 0.1 &    0.3
			\\
						PID                                                                   & 2.4     & 2.4     & 1.6      & 7.0      & 16.0  & 2.4
			     \\
			     			Simulation corrections                                                  & 0.1     & 0.1     & 0.1      & 0.1  & 0.1 &  0.1         \\
			     					Decay model   
		& 1.7    & 2.6     & 4.8      & 4.0  & 5.4 & 0.9
		\\
			Simulated sample size                                                              & 0.2     & 0.2     & 0.2      & 0.3     & 0.5 & 0.1         \\
			$\mathcal{B}(\jpsi\to\mu^{+}\mu^{-})/\mathcal{B}[\lsig\to pK^{-}]$  & 2.3     & 2.3     & 2.3      & 2.3    & 2.3      & 2.3      \\\hline
		Quadratic sum                                                           &10.4 & 7.9 & 10.9 & 12.6 & 22.9 & 8.1\\
  \hline
  $\mathcal{B}(\Lbpkjpsi)$ & 18.8  & 18.8     & 18.8      & 18.8    & 18.8      & 18.8                \\
			\bottomrule
		\end{tabular}
		}
		\label{tab:sys_sum}
        \end{table}

The differential branching fraction measurement is affected by systematic uncertainties in the yield determination and the efficiency estimation. Table~\ref{tab:sys_sum} lists these systematic uncertainties. The total uncertainty is determined from the sum of all contributions in quadrature. The largest uncertainty is related to the uncertainty on the measured mass and width of the \lstar resonances that are fixed in the signal fit model, and is estimated using pseudoexperiments. 
Pseudodata samples are generated according to an alternative model in which the mass and width of the \lstar resonances are varied within their uncertainty~\cite{Hassanvand:2012dn,Sarantsev:2019xxm}, and input values for the Blatt-Weisskopf barrier functions~\cite{Blatt:1952ije} of the \lstar resonances are varied, then fitted with the default model. 

The systematic uncertainty of the \Lb yield determination is evaluated using pseudoexperiments. For both the signal mode \LbLambdastarmumu and the \Lbpkjpsi normalization mode, an alternative model is used where the signal is described by a double-sided crystal ball function~\cite{Skwarnicki:1986xj} and the background by a second-order Chebyshev polynomial function~\cite{smith2010chebyshev}.
Pseudosamples are generated using the alternative model and fitted with the default model, and the observed deviation is assigned as the systematic uncertainty. 
Peaking backgrounds that remain after the vetoes introduced in the selection are neglected in the fit for the determination of the \Lbpkjpsi yield. The main sources of systematic uncertainty are caused by contributions from the \Bskkjpsi and \Bdkpijpsi decays.

The hardware-trigger efficiencies are measured in bins of muon $\pt$ using the \Lbpkjpsi data sample. The effect of an alternative binning scheme on the efficiency ratio is taken as a systematic uncertainty. 
The PID efficiency is determined in bins of the particle momentum and pseudorapidity using calibration data samples. The effects of different binning schemes and different calibration samples
 are evaluated. 
 The sum in quadrature of these effects is taken as the systematic uncertainty arising from the PID efficiency.

The systematic uncertainty associated with the simulation corrections is determined by using alternative binning schemes to account for the finite statistics of the control modes. The corresponding deviation is taken as the systematic uncertainty.
 The systematic uncertainty associated with the decay model used in simulation is estimated by taking the difference in efficiency between the phase-space model and the model given in Ref~\cite{Descotes-Genon:2019dbw}.
In addition, the systematic uncertainties due to the limited size of the simulated sample and precision of the $\jpsi\to\mumu$ and $\lsig\to\proton\Km$ branching fractions are also taken into account.

\begin{table}
\centering
  \caption{
    Signal yields and the absolute differential branching fraction, in bins of $\qsq$, for the \LbLambdastarmumu decay. The first uncertainty is statistical, the second systematic, and the third due to the uncertainty on the $\Lbpkjpsi$ and $\jpsi\to\mu^+\mu^-$ branching fractions. 
    \label{tab:bfresults}
  }
\label{tab:results}
\renewcommand*{\arraystretch}{1.25}
\begin{tabular}{ccc}
\toprule
\noalign{\smallskip}
   \qsq interval $(\!\gevgevcccc)$   & $N_{\Lambdastarmumu}$ & $d\mathcal{B}[\LbLambdastarmumu]/d\qsq$ ($10^{-8}\gev^{-2}c^{4}$)\\
    \noalign{\smallskip}
    \midrule
    $0.1\text{--}3.0$    & $\phantom{0}96 \pm 18$ & $1.89\pm 0.35 \pm 0.19 \pm 0.36$ \\
    $3.0\text{--}6.0$        & $138 \pm 18$ & $2.42\pm 0.32 \pm 0.17 \pm 0.45$\\
    $6.0\text{--}8.0$        &  $\phantom{0}65 \pm 14$ & $1.58\pm0.36\pm0.16\pm0.30$\\
    $11.0\text{--}12.5$        & $\phantom{0}59 \pm 14$ & $2.07\pm0.47\pm0.26\pm0.39$\\
    $15.0\text{--}17.0$    & $\phantom{0}12 \pm \phantom{0}5$ & $0.57\pm0.24\pm0.13\pm0.11$\\
\midrule
 $1.1\text{--}6.0$        & $175 \pm 21$ & $1.95\pm0.23\pm0.16\pm0.37$\\
        \bottomrule
\end{tabular}
\end{table}

\begin{figure}[b]
  \centering
  \includegraphics[width=0.7\linewidth]{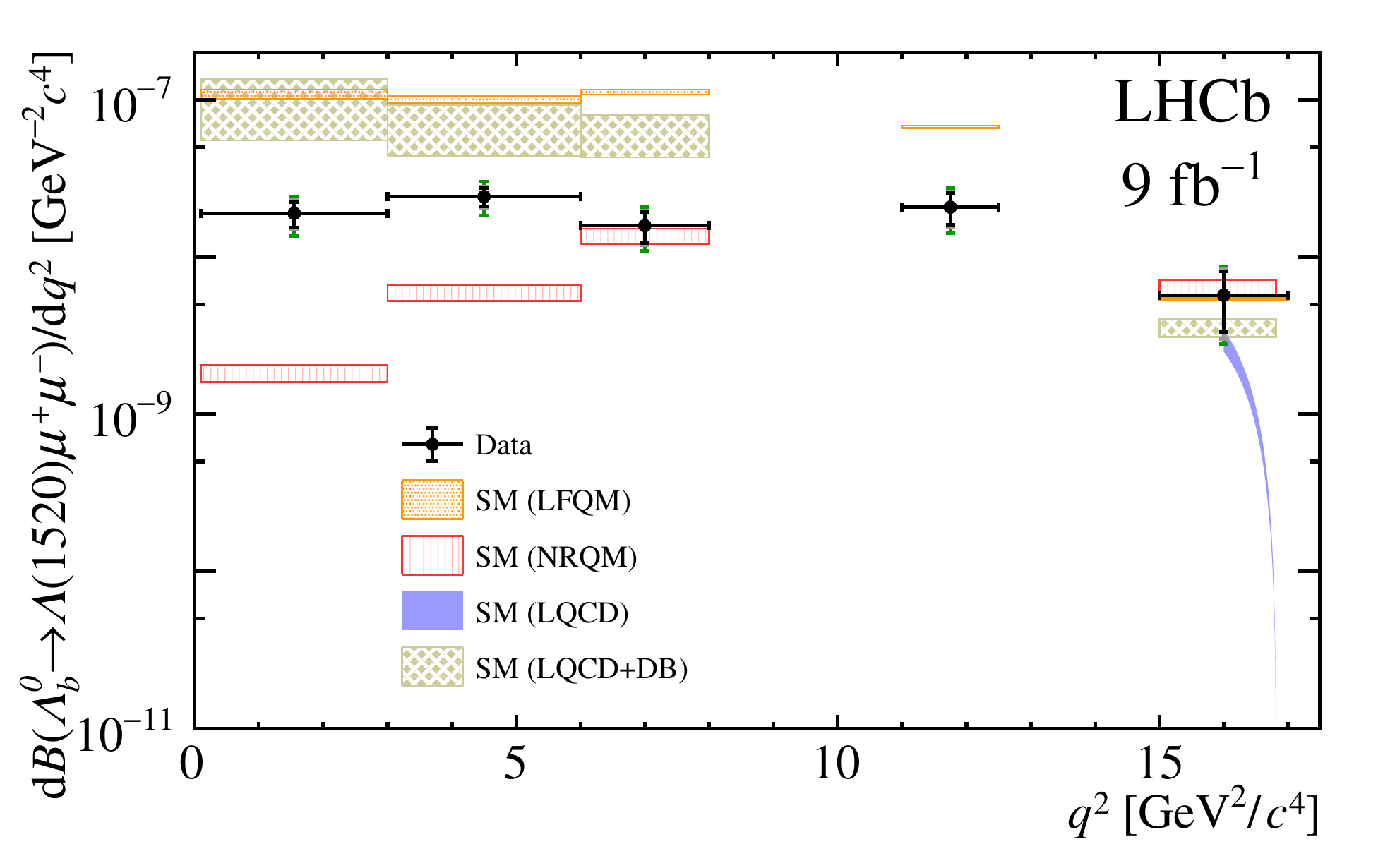}

  \caption{Differential branching fraction of the \LbLambdastarmumu decay in intervals of \qsq. The error bars in black, gray and green represent the measured results with statistical, systematic and $\mathcal{B}(\Lbpkjpsi)$ uncertainties taken into account. Also shown are the SM predictions
  using the form factors calculated with the nonrelativistic quark model (NRQM)~\cite{Descotes-Genon:2019dbw}, light-front quark model (LFQM)~\cite{Li:2022nim}, joint lattice QCD and dispersive bound (LQCD+DB)~\cite{Amhis:2022vcd} and lattice QCD (LQCD)~\cite{Meinel:2021mdj}. Note that the LQCD prediction is only available for $\qsq$ above $16\gevgevcccc$, and the trend instead of a rate average is shown. These theoretical predictions as well as evaluation of the corresponding uncertainties should be revisited given their significant differences. }
  \label{fig:bf}
\end{figure}

The differential branching fraction of the \LbLambdastarmumu decay in intervals of $\qsq$ is reported in Table~\ref{tab:bfresults}, and is shown in Fig.~\ref{fig:bf}.
The SM prediction from Ref.~\cite{Descotes-Genon:2019dbw}, for which only the form factor uncertainties are considered, and 
the SM prediction from Refs.~\cite{Li:2022nim} and~\cite{Amhis:2022vcd}, are also shown. It is impossible to make a firm statement about the level of agreement between the experimental measurement and the theoretical predictions due to the significant variation among the different theoretical predictions.

In summary, the first measurement of the branching fraction of the rare decay $\LbLambdastarmumu$ is presented using LHCb data corresponding to a total integrated luminosity of $9\invfb$.
The data are compared to several predictions within SM. In the highest $\qsq$ interval, $\qsq >$15.0\gevgevcccc, where the predictions have the smallest model dependence,
they are consistent with the data.
In the low-$\qsq$ region, it is not possible to make a statement about the agreement between this experimental result and the predictions since the variation between the different SM predictions is much larger than their quoted uncertainties. 

\clearpage

\section*{Supplemental material}
\label{sec:Supp}

\begin{figure}[b]
  \centering
  \includegraphics[width=0.49\linewidth]{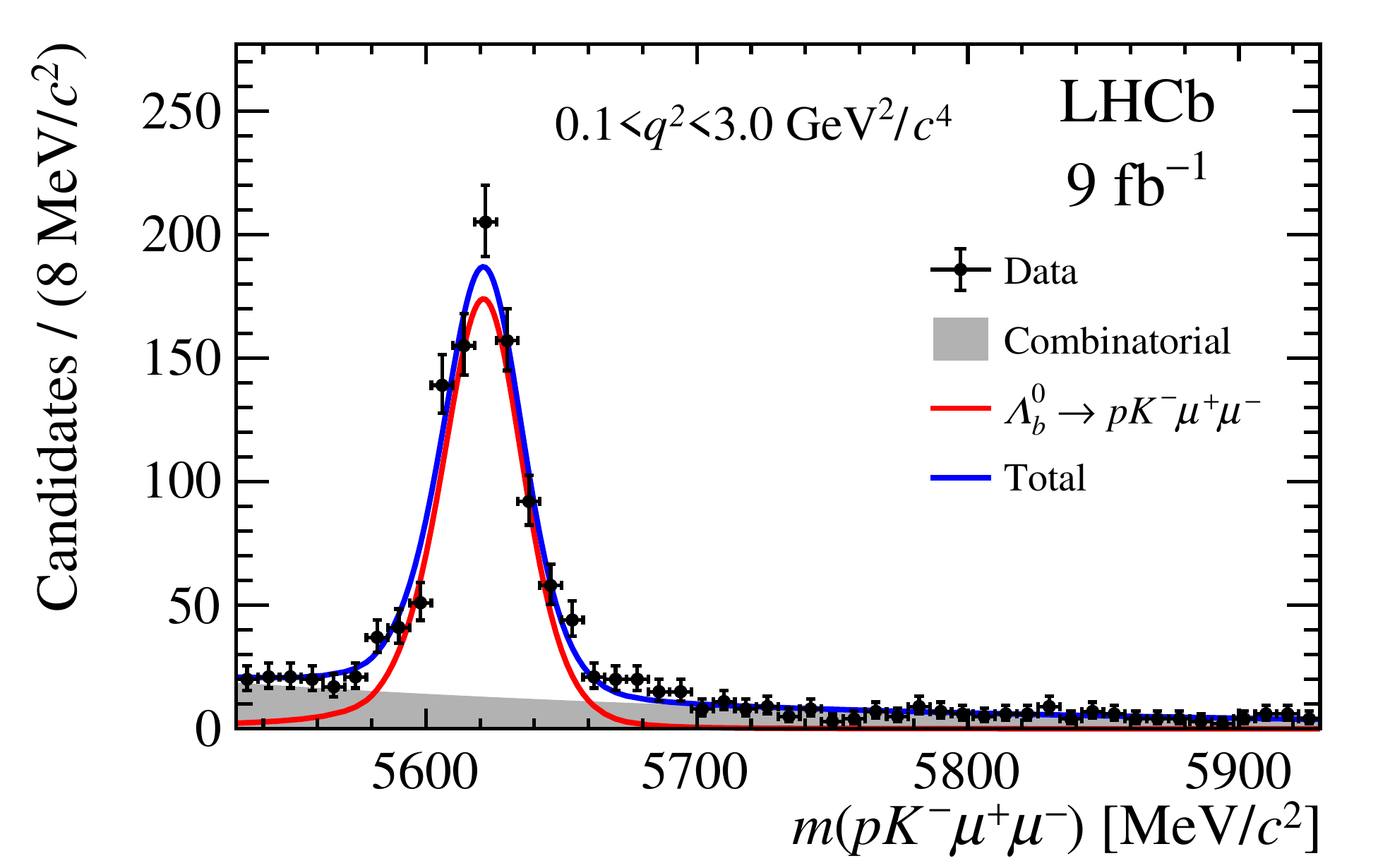}
  \includegraphics[width=0.49\linewidth]{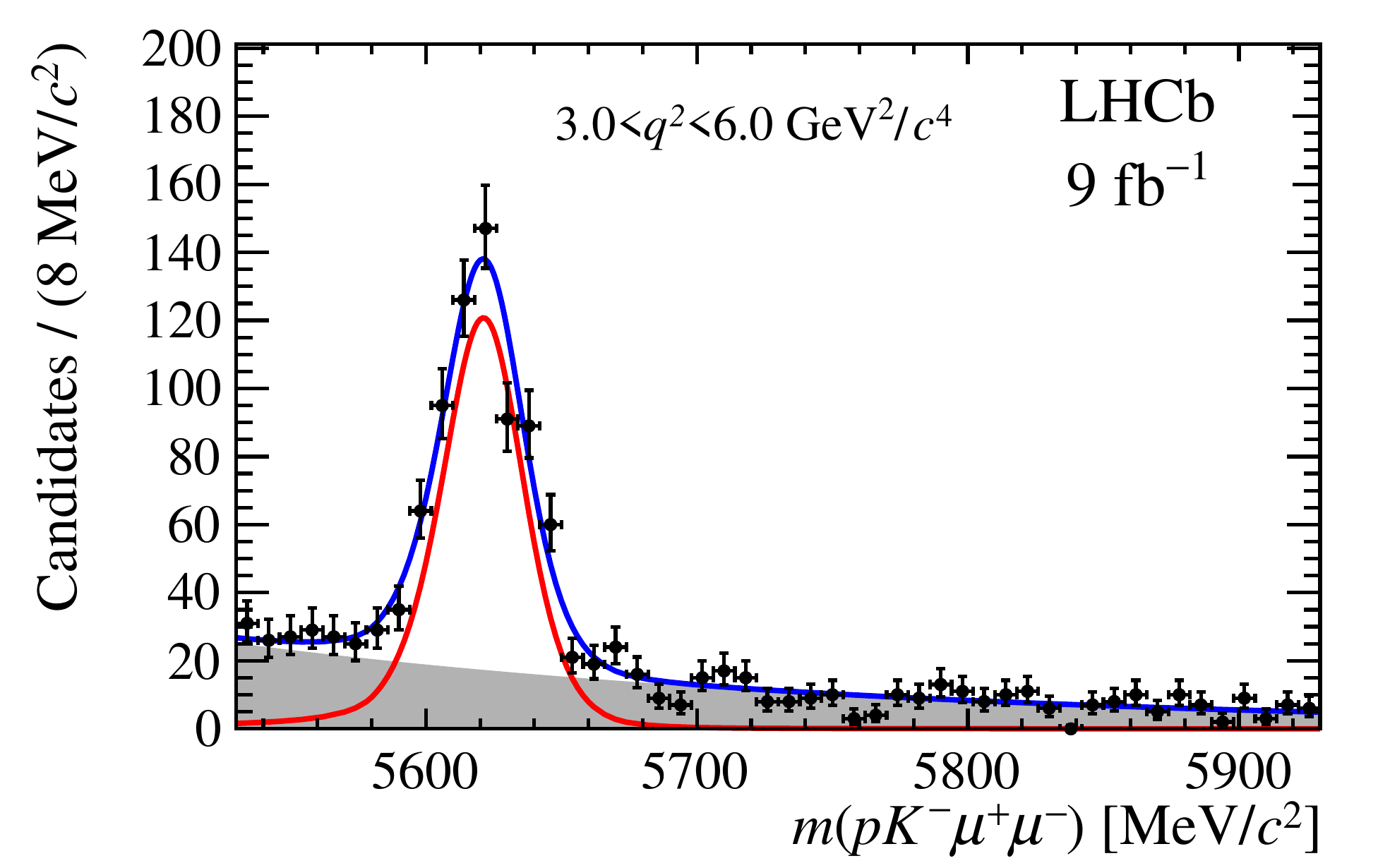}

  \includegraphics[width=0.49\linewidth]{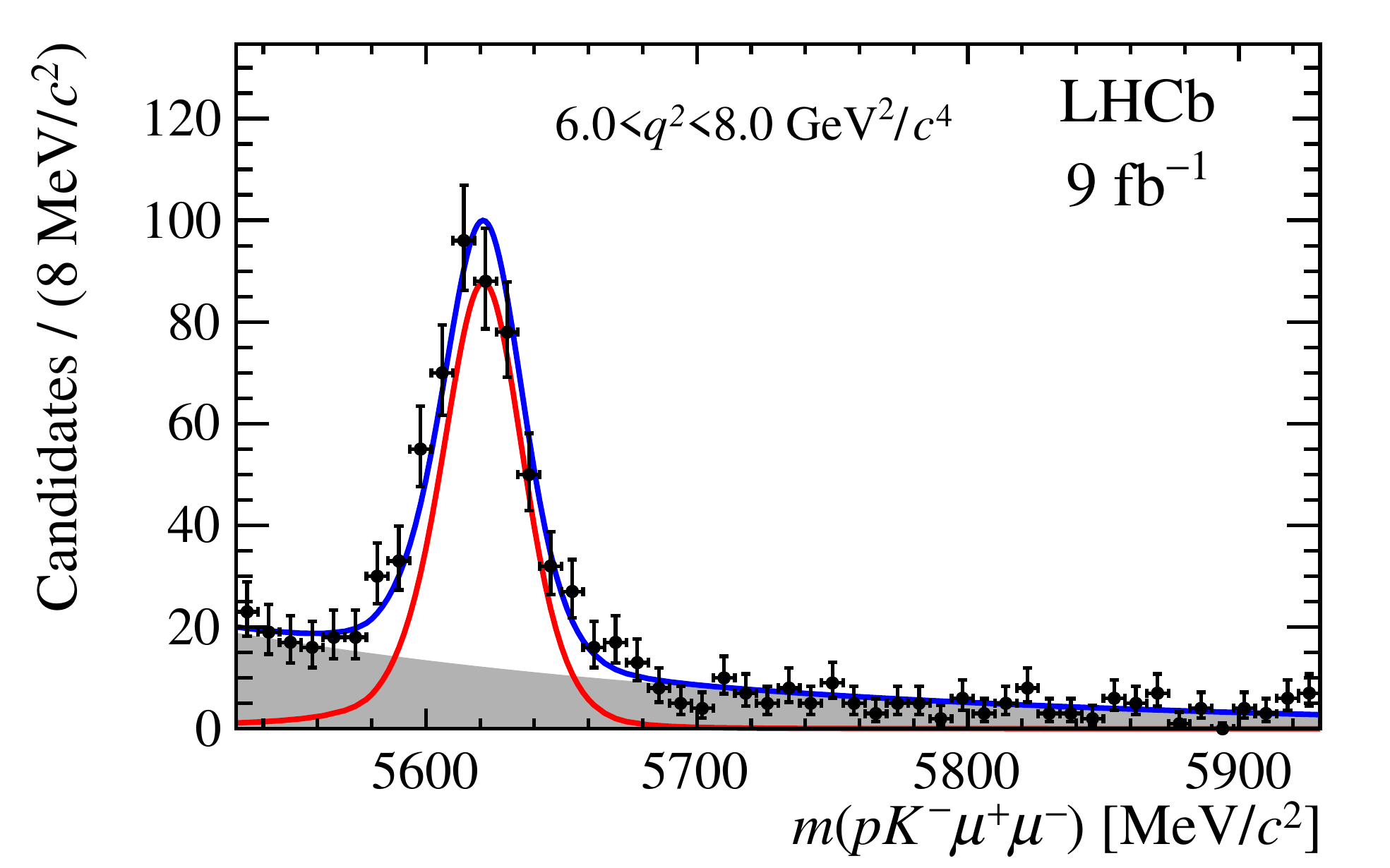}
  \includegraphics[width=0.49\linewidth]{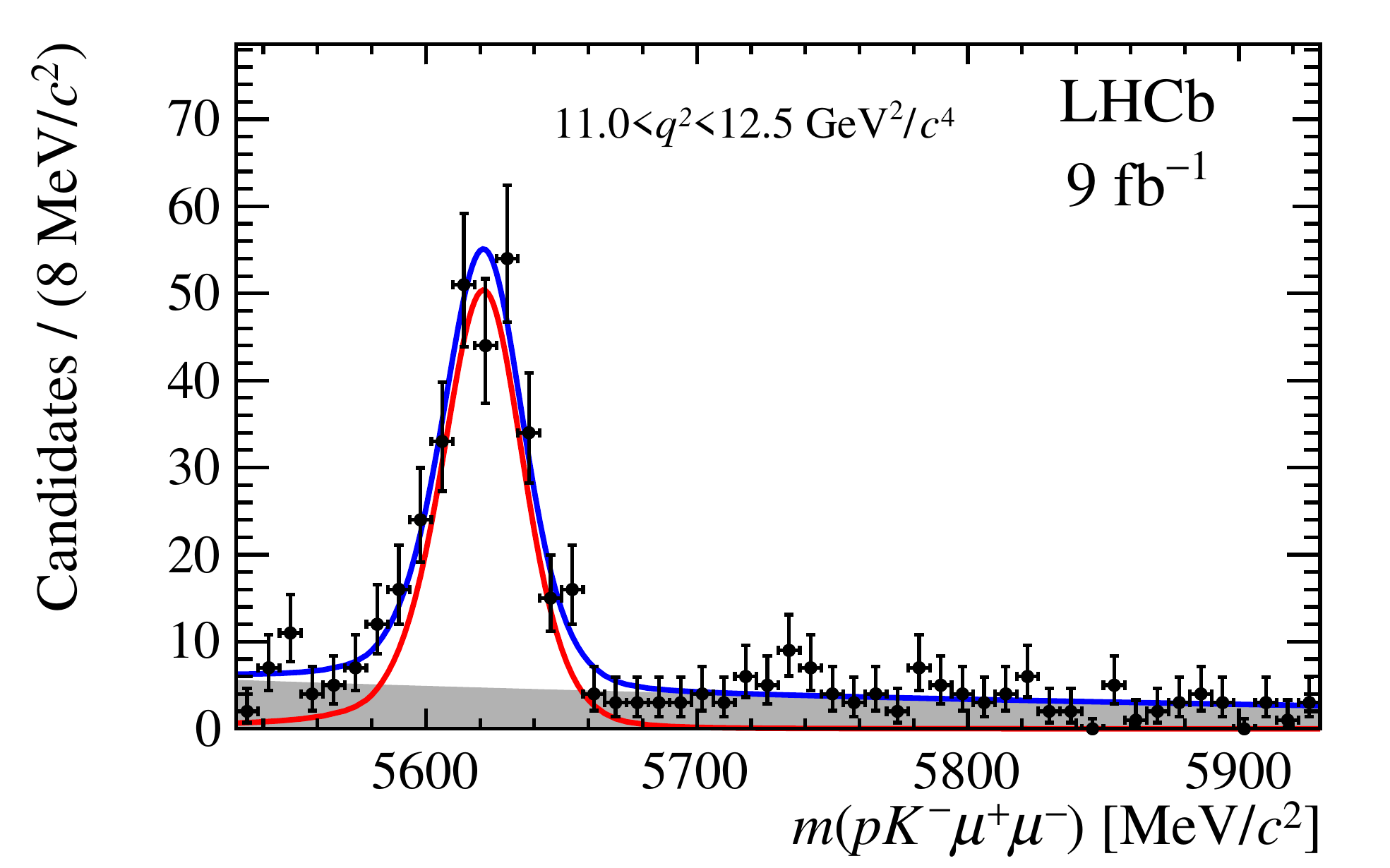}

  \includegraphics[width=0.49\linewidth]{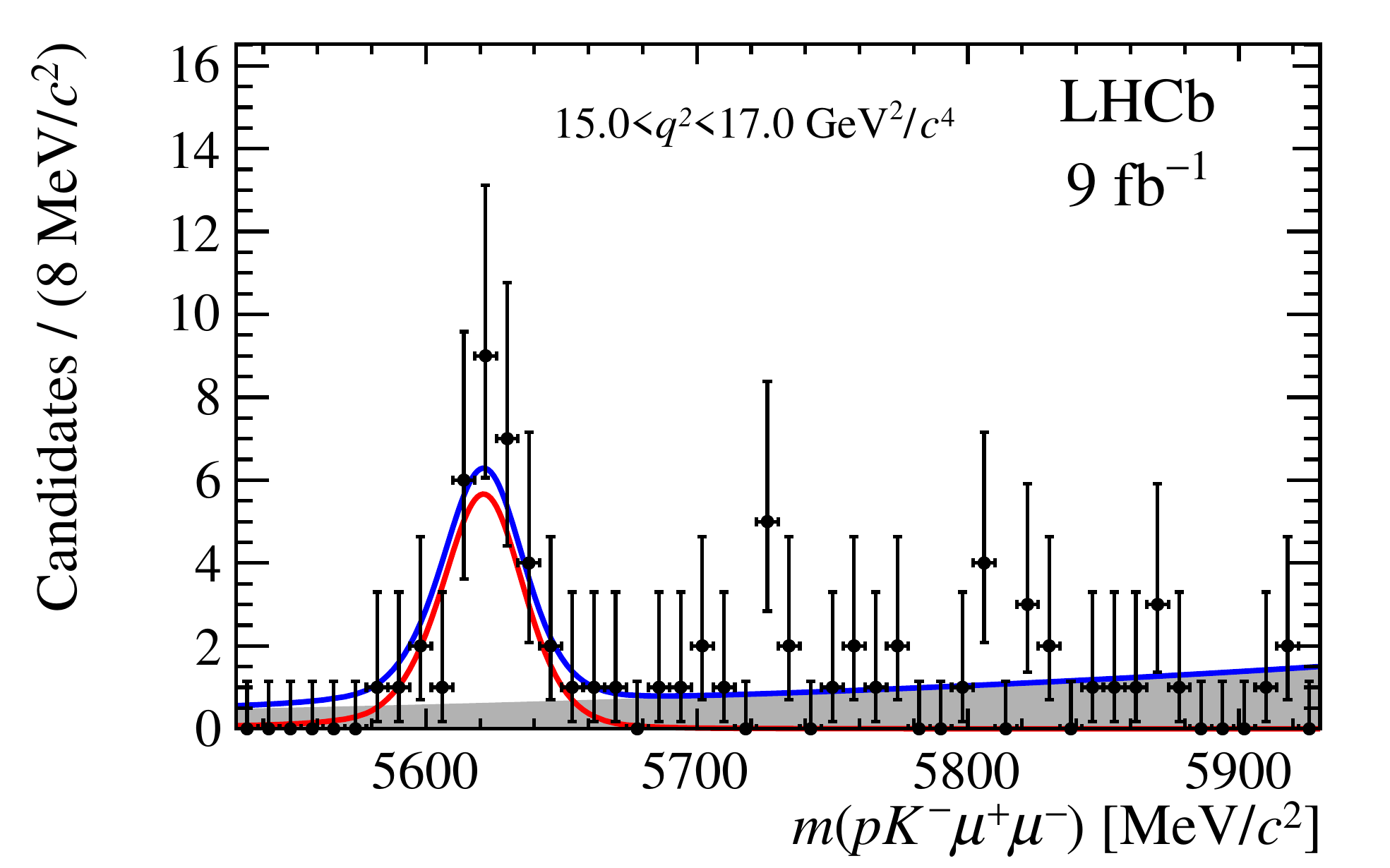}
  \includegraphics[width=0.49\linewidth]{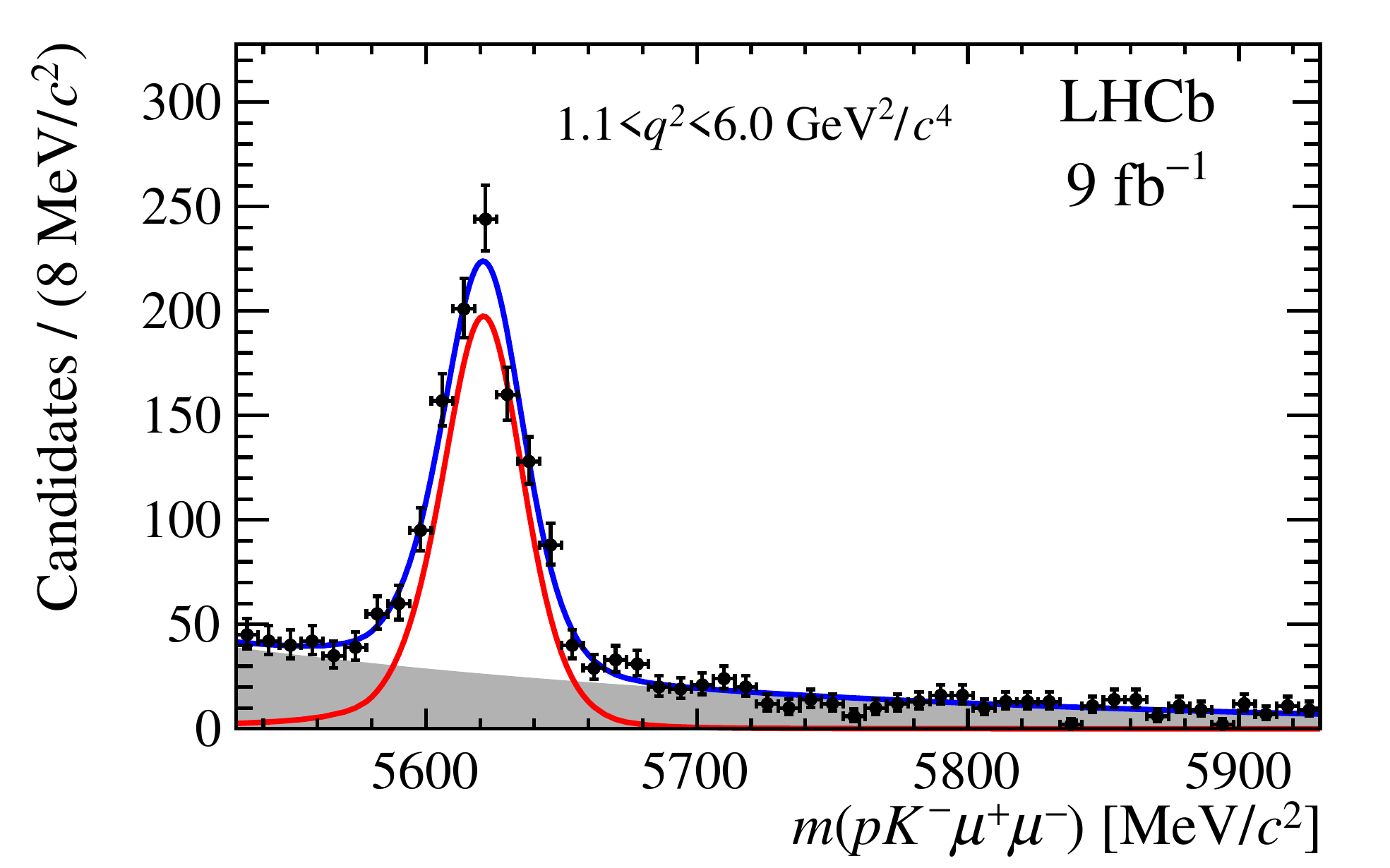}

  \caption{Mass distributions of the \Lbpkmumu signal candidates in different \qsq intervals. The data are overlaid with the fit projections.}
  \label{fig:lbfit_q2}
\end{figure}

\begin{figure}
  \centering
  \includegraphics[width=0.49\linewidth]{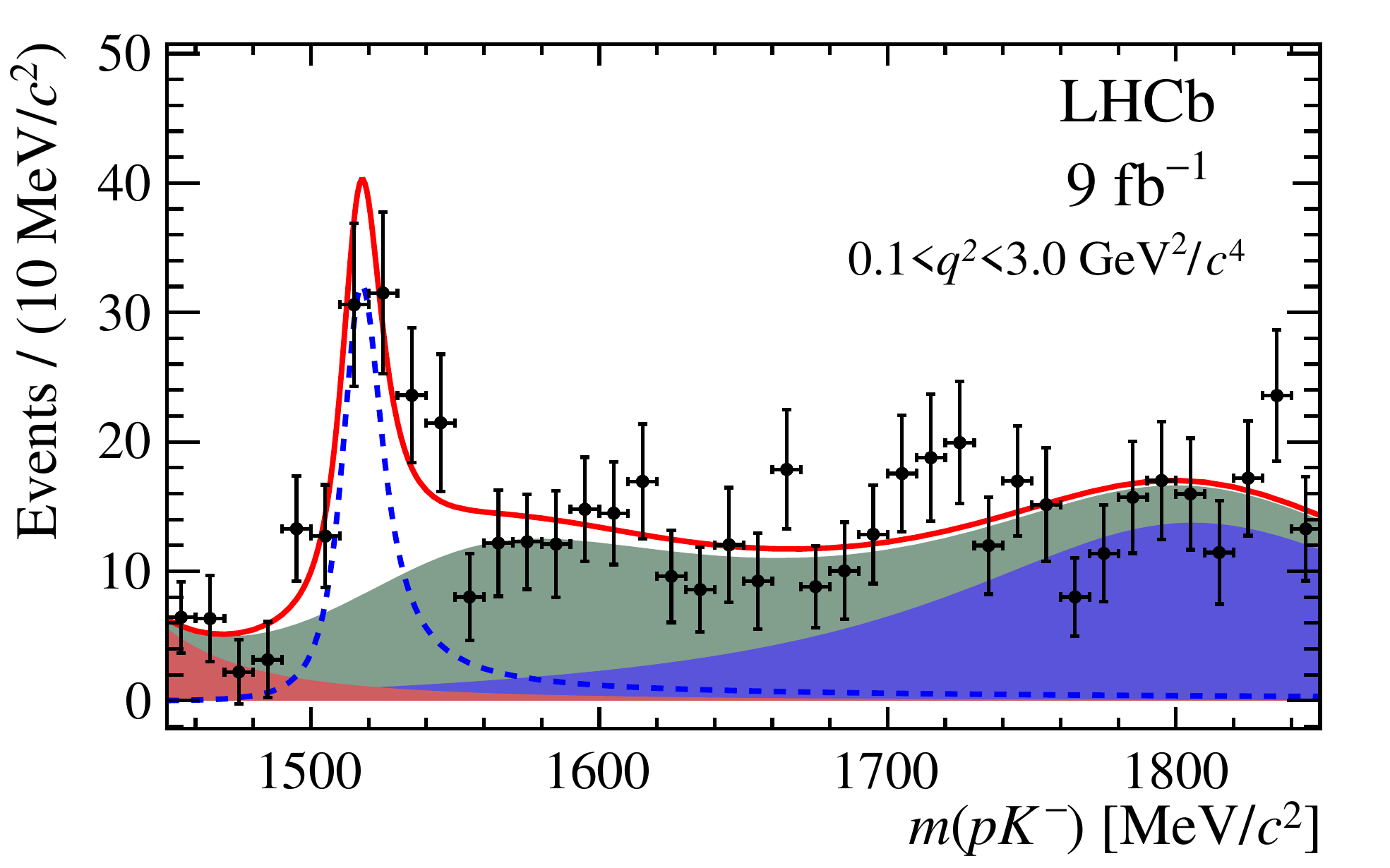}
  \includegraphics[width=0.49\linewidth]{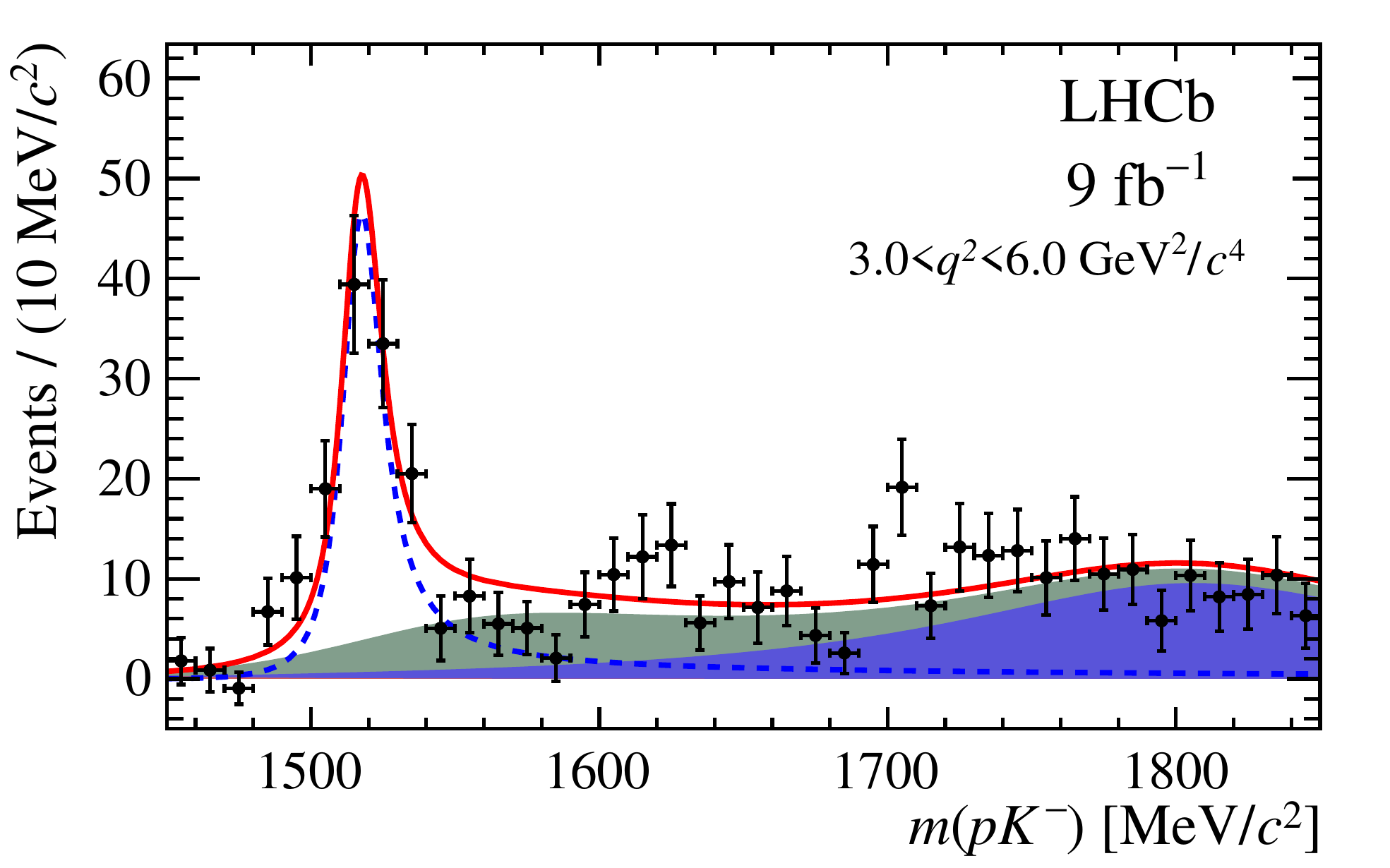}

  \includegraphics[width=0.49\linewidth]{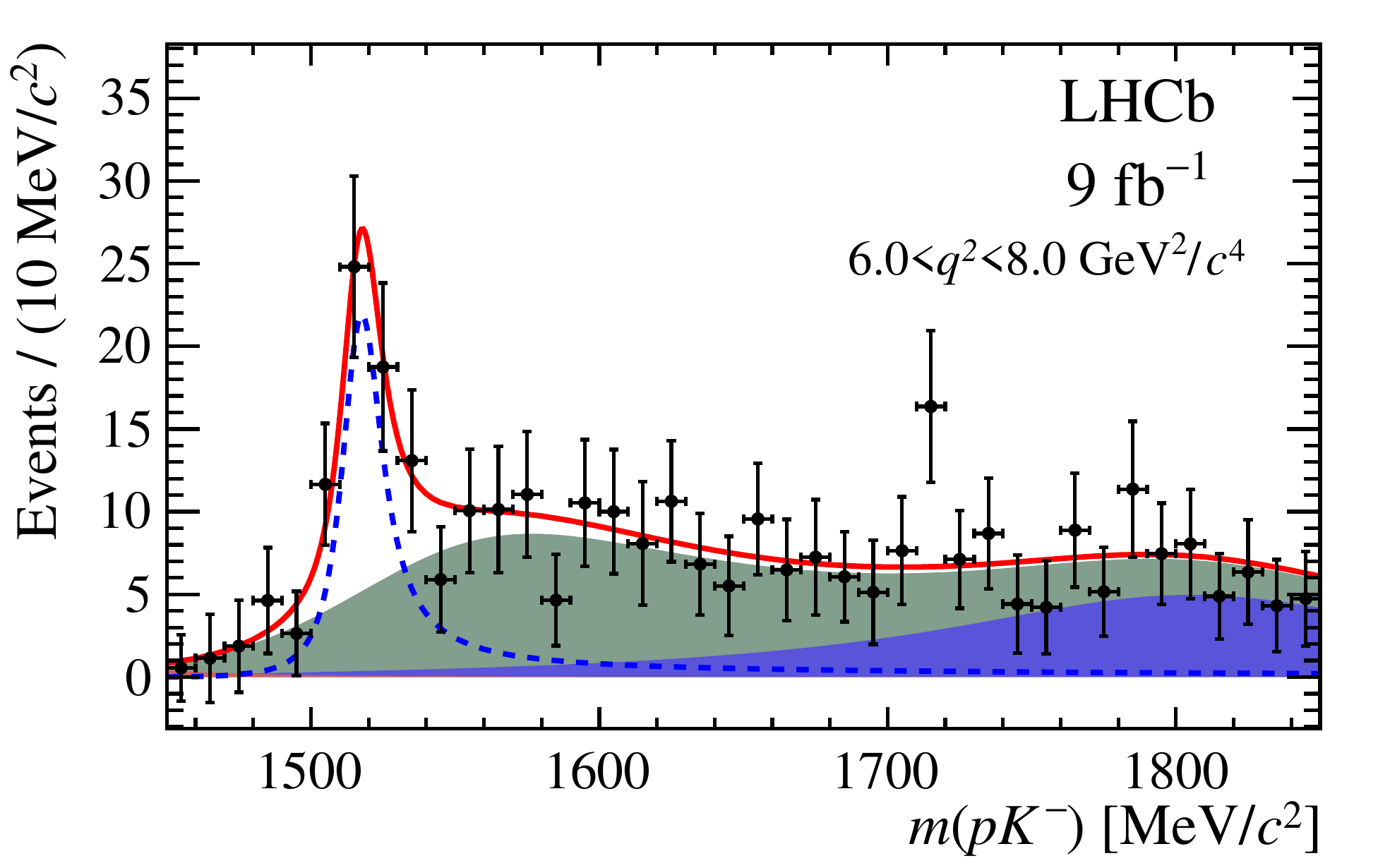}
  \includegraphics[width=0.49\linewidth]{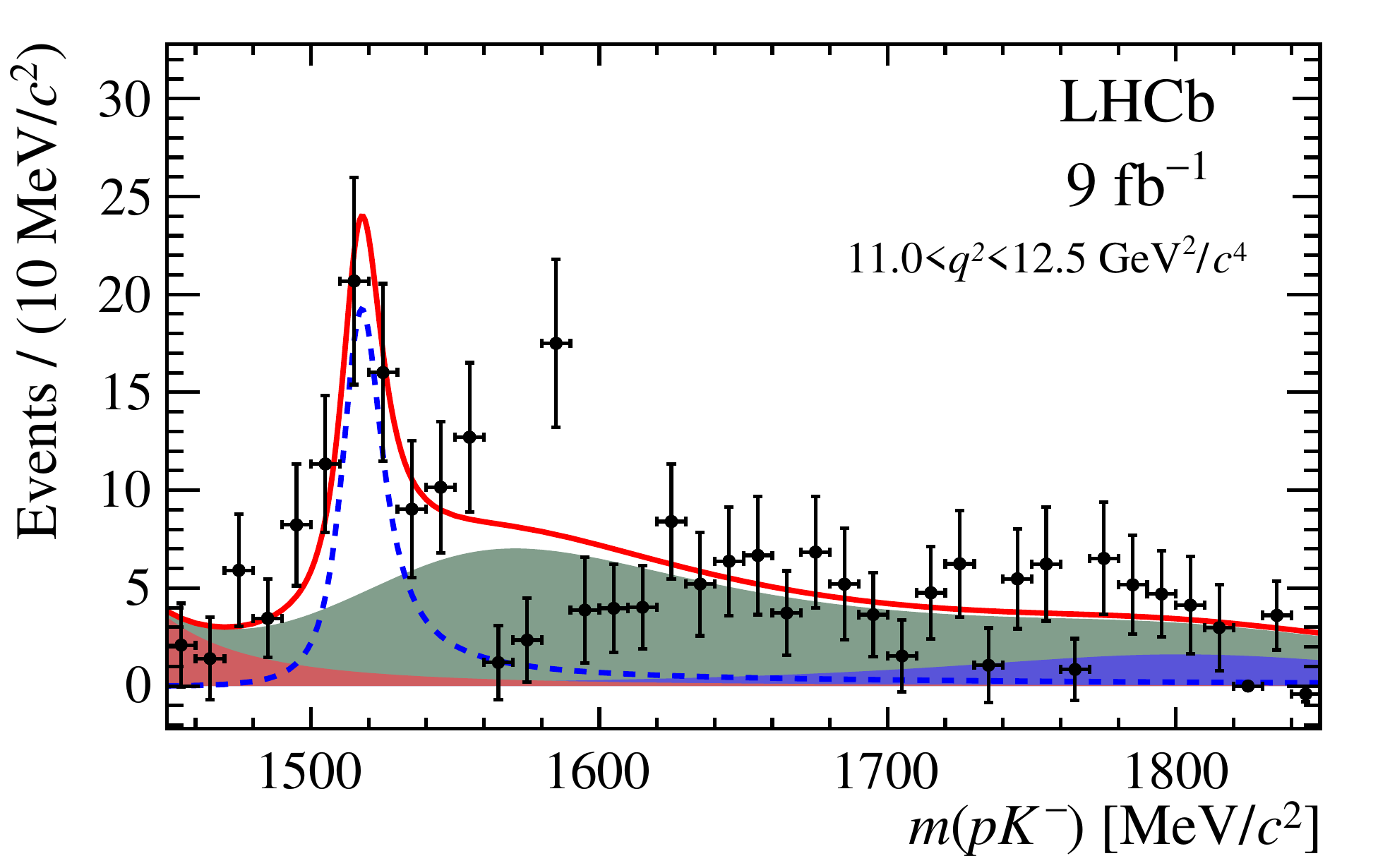}

  \includegraphics[width=0.49\linewidth]{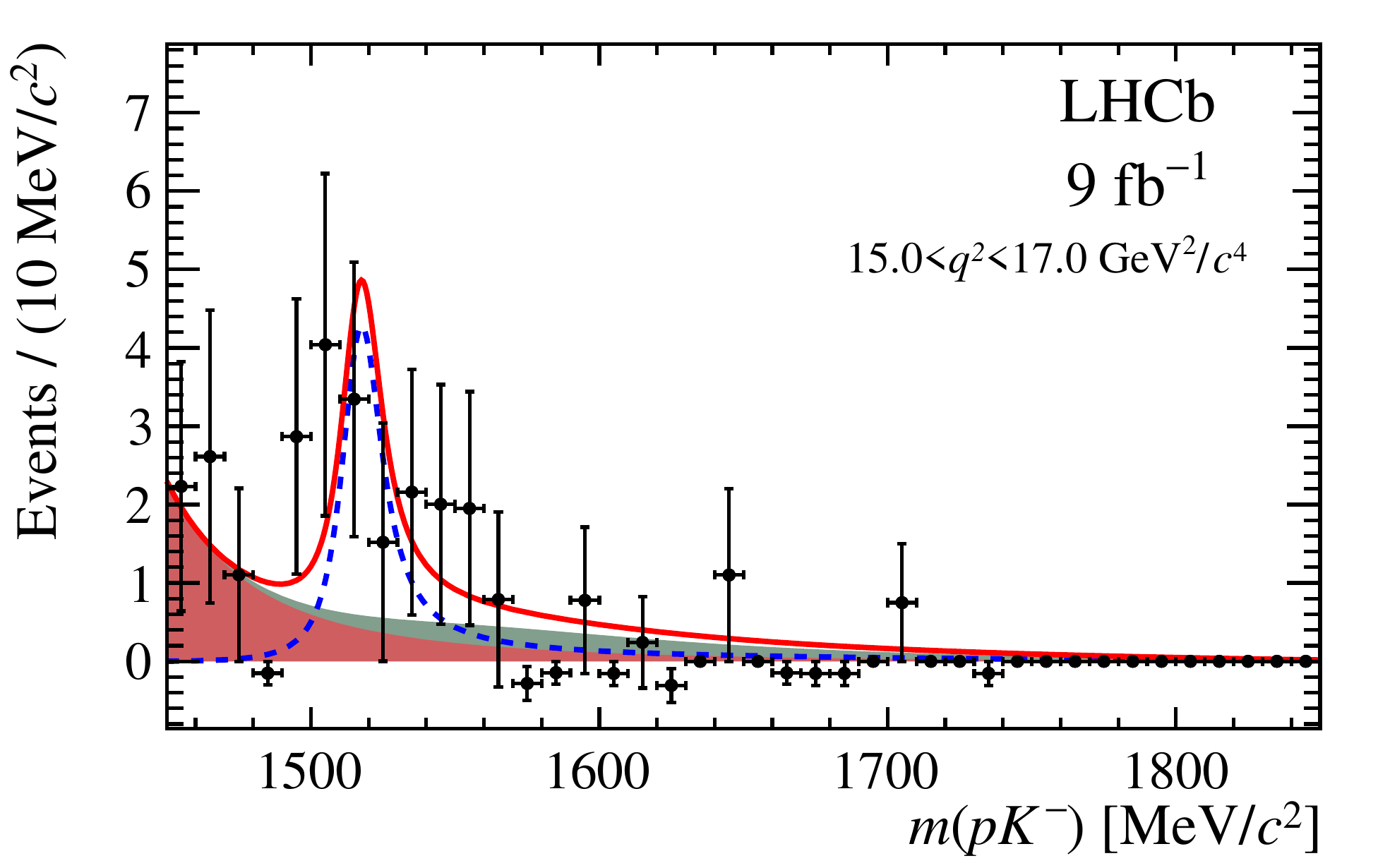}
  \includegraphics[width=0.49\linewidth]{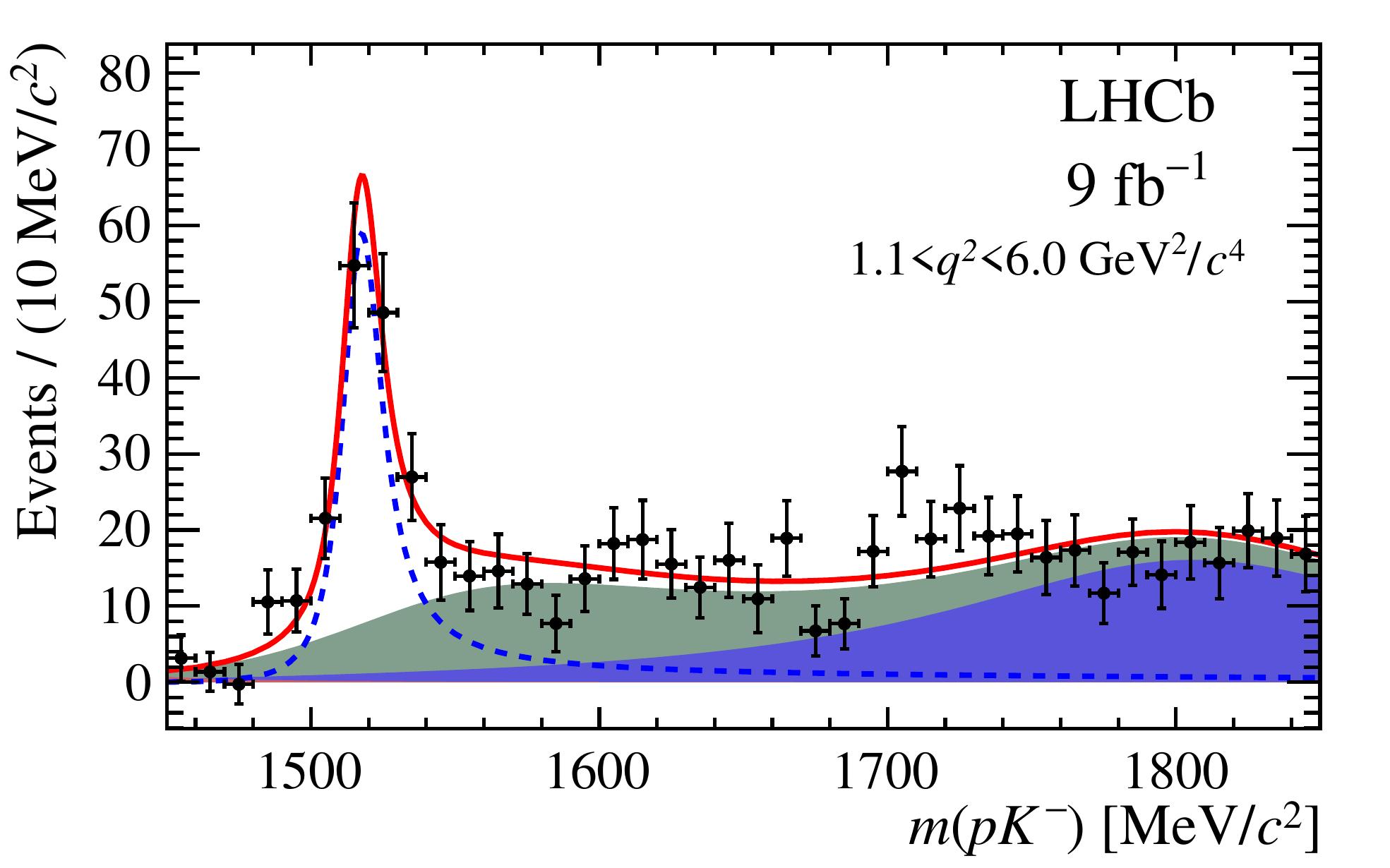}

  \caption{Distribution of $\proton\Km$ mass, after background subtraction,  for the \LbLambdastarmumu signal candidates in intervals of \qsq. Distributions are overlaid with the fit projections.}
  \label{fig:xmfit_q2}
\end{figure}

\begin{figure}
  \centering
  \includegraphics[width=0.49\linewidth]{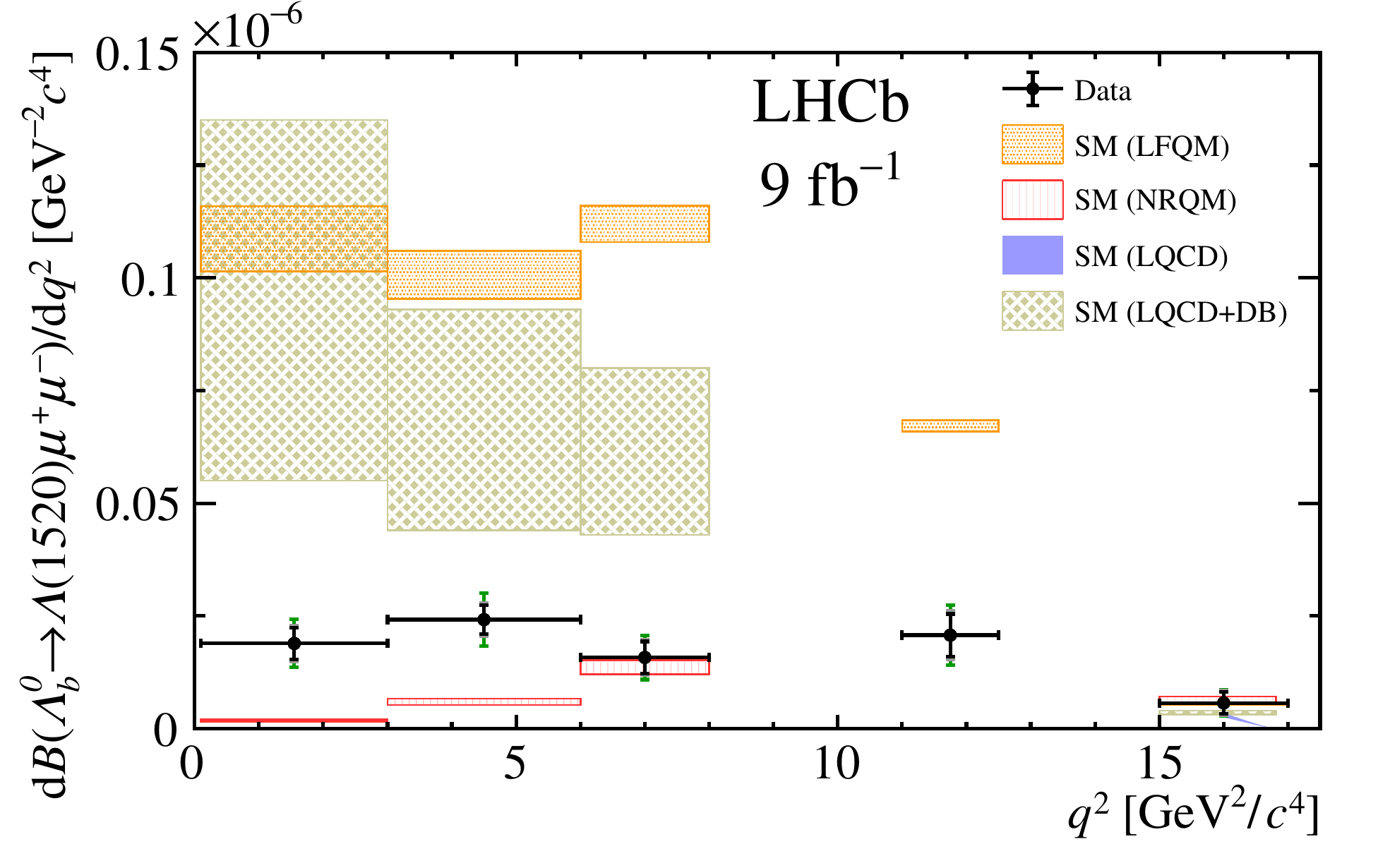}
  \includegraphics[width=0.49\linewidth]{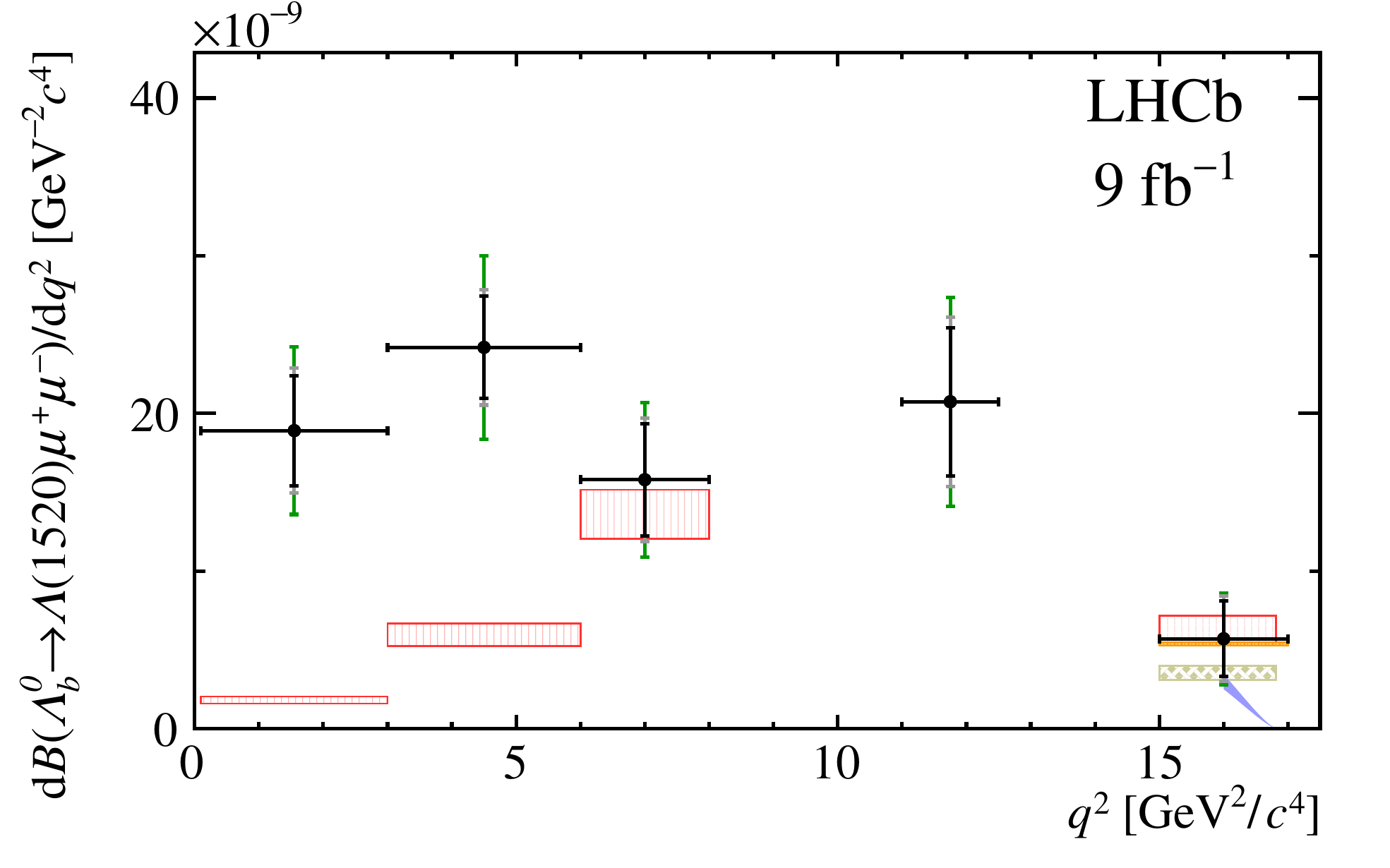}
  \includegraphics[width=0.49\linewidth]{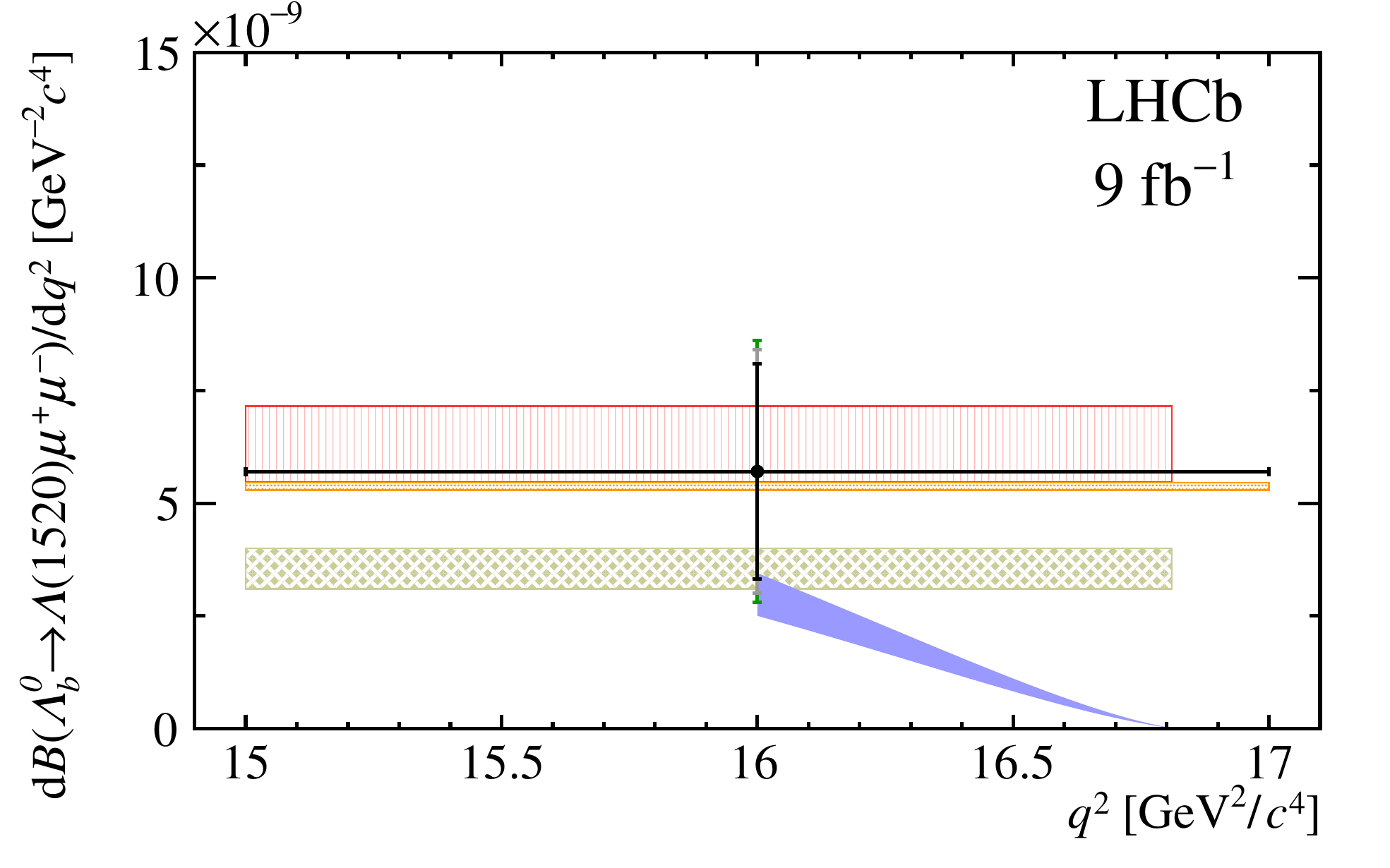}

  \caption{
  Comparison between data and SM predictions. Note that the LQCD prediction is only available for $\qsq$ above $16\gevgevcccc$, and the trend instead of a rate average is shown.
  }
  \label{fig:bf_zoom}
\end{figure}

Figure~\ref{fig:lbfit_q2} shows the mass distributions of the \Lbpkmumu signal candidates in the five \qsq intervals, and integrated over the first two intervals. The total fit projection (blue line) is overlaid on the data, along with the signal component (red line) and background component describing combinatorial background (grey shadow).
Figure~\ref{fig:xmfit_q2} shows the $\proton\Km$ mass for \LbLambdastarmumu signal candidates in different \qsq bins. The total fit projection (red line) is overlaid on the data along with projections of individual fit components describing: the signal (blue dashed line) and the \llow (red), \lmid (green) and \lhigh (purple) components.
Figure~\ref{fig:bf_zoom} shows the comparison between data and SM predictions in the zoomed ranges of Fig.~\ref{fig:bf}.

\clearpage

\section*{Acknowledgements}
%
%
\noindent We express our gratitude to our colleagues in the CERN
accelerator departments for the excellent performance of the LHC. We
thank the technical and administrative staff at the LHCb
institutes.
We acknowledge support from CERN and from the national agencies:
CAPES, CNPq, FAPERJ and FINEP (Brazil); 
MOST and NSFC (China); 
CNRS/IN2P3 (France); 
BMBF, DFG and MPG (Germany); 
INFN (Italy); 
NWO (Netherlands); 
MNiSW and NCN (Poland); 
MEN/IFA (Romania); 
MICINN (Spain); 
SNSF and SER (Switzerland); 
NASU (Ukraine); 
STFC (United Kingdom); 
DOE NP and NSF (USA).
We acknowledge the computing resources that are provided by CERN, IN2P3
(France), KIT and DESY (Germany), INFN (Italy), SURF (Netherlands),
PIC (Spain), GridPP (United Kingdom), 
CSCS (Switzerland), IFIN-HH (Romania), CBPF (Brazil),
Polish WLCG  (Poland) and NERSC (USA).
We are indebted to the communities behind the multiple open-source
software packages on which we depend.
Individual groups or members have received support from
ARC and ARDC (Australia);
Minciencias (Colombia);
AvH Foundation (Germany);
EPLANET, Marie Sk\l{}odowska-Curie Actions and ERC (European Union);
A*MIDEX, ANR, IPhU and Labex P2IO, and R\'{e}gion Auvergne-Rh\^{o}ne-Alpes (France);
Key Research Program of Frontier Sciences of CAS, CAS PIFI, CAS CCEPP, 
Fundamental Research Funds for the Central Universities, 
and Sci. \& Tech. Program of Guangzhou (China);
GVA, XuntaGal, GENCAT and Prog.~Atracci\'on Talento, CM (Spain);
SRC (Sweden);
the Leverhulme Trust, the Royal Society
 and UKRI (United Kingdom).

\addcontentsline{toc}{section}{References}
\ifx\mcitethebibliography\mciteundefinedmacro
\PackageError{LHCb.bst}{mciteplus.sty has not been loaded}
{This bibstyle requires the use of the mciteplus package.}\fi
\providecommand{\href}[2]{#2}

\newpage
\centerline
{\large\bf LHCb collaboration}
\begin
{flushleft}
\small
R.~Aaij$^{32}$\lhcborcid{0000-0003-0533-1952},
A.S.W.~Abdelmotteleb$^{50}$\lhcborcid{0000-0001-7905-0542},
C.~Abellan~Beteta$^{44}$,
F.~Abudin{\'e}n$^{50}$\lhcborcid{0000-0002-6737-3528},
T.~Ackernley$^{54}$\lhcborcid{0000-0002-5951-3498},
B.~Adeva$^{40}$\lhcborcid{0000-0001-9756-3712},
M.~Adinolfi$^{48}$\lhcborcid{0000-0002-1326-1264},
P.~Adlarson$^{77}$\lhcborcid{0000-0001-6280-3851},
H.~Afsharnia$^{9}$,
C.~Agapopoulou$^{13}$\lhcborcid{0000-0002-2368-0147}
% ...

{\footnotesize \it

% ...
$^{9}$Universit{\'e} Clermont Auvergne, CNRS/IN2P3, LPC, Clermont-Ferrand, France\\
% ...
$^{13}$LPNHE, Sorbonne Universit{\'e}, Paris Diderot Sorbonne Paris Cit{\'e}, CNRS/IN2P3, Paris, France\\
% ...
$^{32}$Nikhef National Institute for Subatomic Physics, Amsterdam, Netherlands\\
% ...
$^{40}$Instituto Galego de F{\'\i}sica de Altas Enerx{\'\i}as (IGFAE), Universidade de Santiago de Compostela, Santiago de Compostela, Spain\\
% ...
$^{44}$Physik-Institut, Universit{\"a}t Z{\"u}rich, Z{\"u}rich, Switzerland\\
% ...
$^{48}$H.H. Wills Physics Laboratory, University of Bristol, Bristol, United Kingdom\\
% ...
$^{50}$Department of Physics, University of Warwick, Coventry, United Kingdom\\
% ...
$^{53}$School of Physics and Astronomy, University of Glasgow, Glasgow, United Kingdom\\
$^{54}$Oliver Lodge Laboratory, University of Liverpool, Liverpool, United Kingdom\\
% ...
$^{77}$Department of Physics and Astronomy, Uppsala University, Uppsala, Sweden, associated to $^{53}$\\
% ...
}
\end{flushleft}


\begin{mcitethebibliography}{10}
\mciteSetBstSublistMode{n}
\mciteSetBstMaxWidthForm{subitem}{\alph{mcitesubitemcount})}
\mciteSetBstSublistLabelBeginEnd{\mcitemaxwidthsubitemform\space}
{\relax}{\relax}

\bibitem{Das:2018sms}
D.~Das, \ifthenelse{\boolean{articletitles}}{\emph{{Model independent new
  physics analysis in $\PLambda_b\to\PLambda\mu^+\mu^-$ decay}},
  }{}\href{https://doi.org/10.1140/epjc/s10052-018-5731-2}{Eur.\ Phys.\ J.\
  \textbf{C78} (2018) 230},
  \href{http://arxiv.org/abs/1802.09404}{{\normalfont\ttfamily
  arXiv:1802.09404}}\relax
\mciteBstWouldAddEndPuncttrue
\mciteSetBstMidEndSepPunct{\mcitedefaultmidpunct}
{\mcitedefaultendpunct}{\mcitedefaultseppunct}\relax
\EndOfBibitem
\bibitem{Langacker:2000ju}
P.~Langacker and M.~Plumacher,
  \ifthenelse{\boolean{articletitles}}{\emph{{Flavor changing effects in
  theories with a heavy $Z^\prime$ boson with family nonuniversal couplings}},
  }{}\href{https://doi.org/10.1103/PhysRevD.62.013006}{Phys.\ Rev.\
  \textbf{D62} (2000) 013006},
  \href{http://arxiv.org/abs/hep-ph/0001204}{{\normalfont\ttfamily
  arXiv:hep-ph/0001204}}\relax
\mciteBstWouldAddEndPuncttrue
\mciteSetBstMidEndSepPunct{\mcitedefaultmidpunct}
{\mcitedefaultendpunct}{\mcitedefaultseppunct}\relax
\EndOfBibitem
\bibitem{Biswas:2022fyb}
S.~Biswas, S.~Mahata, B.~P. Nayak, and S.~Sahoo,
  \ifthenelse{\boolean{articletitles}}{\emph{{Imprints of new physics in
  $\PLambda_b \rightarrow \PLambda^* \ell^+ \ell^-$ decay in nonuniversal
  $Z^{\prime}$ model}},
  }{}\href{https://doi.org/10.1142/S0217751X22500440}{Int.\ J.\ Mod.\ Phys.\
  \textbf{A37} (2022) 2250044}\relax
\mciteBstWouldAddEndPuncttrue
\mciteSetBstMidEndSepPunct{\mcitedefaultmidpunct}
{\mcitedefaultendpunct}{\mcitedefaultseppunct}\relax
\EndOfBibitem
\bibitem{LHCb-PAPER-2013-043}
LHCb collaboration, R.~Aaij {\em et~al.},
  \ifthenelse{\boolean{articletitles}}{\emph{{Measurement of the \CP asymmetry
  in \mbox{\decay{\Bp}{\Kp\mumu}} decays}},
  }{}\href{https://doi.org/10.1103/PhysRevLett.111.151801}{Phys.\ Rev.\ Lett.\
  \textbf{111} (2013) 151801},
  \href{http://arxiv.org/abs/1308.1340}{{\normalfont\ttfamily
  arXiv:1308.1340}}\relax
\mciteBstWouldAddEndPuncttrue
\mciteSetBstMidEndSepPunct{\mcitedefaultmidpunct}
{\mcitedefaultendpunct}{\mcitedefaultseppunct}\relax
\EndOfBibitem
\bibitem{LHCb-PAPER-2014-006}
LHCb collaboration, R.~Aaij {\em et~al.},
  \ifthenelse{\boolean{articletitles}}{\emph{{Differential branching fractions
  and isospin asymmetries of \mbox{\decay{\B}{K^{(*)}\mumu}} decays}},
  }{}\href{https://doi.org/10.1007/JHEP06(2014)133}{JHEP \textbf{06} (2014)
  133}, \href{http://arxiv.org/abs/1403.8044}{{\normalfont\ttfamily
  arXiv:1403.8044}}\relax
\mciteBstWouldAddEndPuncttrue
\mciteSetBstMidEndSepPunct{\mcitedefaultmidpunct}
{\mcitedefaultendpunct}{\mcitedefaultseppunct}\relax
\EndOfBibitem
\bibitem{LHCb-PAPER-2016-012}
LHCb collaboration, R.~Aaij {\em et~al.},
  \ifthenelse{\boolean{articletitles}}{\emph{{Measurements of the S-wave
  fraction in \mbox{\decay{\Bz}{\Kp\pim\mumu}} decays and the
  \mbox{\decay{\Bz}{\Kstar(892)^0\mumu}} differential branching fraction}},
  }{}\href{https://doi.org/10.1007/JHEP11(2016)047}{JHEP \textbf{11} (2016)
  047}, Erratum \href{https://doi.org/10.1007/JHEP04(2017)142}{ibid.\
  \textbf{04} (2017) 142},
  \href{http://arxiv.org/abs/1606.04731}{{\normalfont\ttfamily
  arXiv:1606.04731}}\relax
\mciteBstWouldAddEndPuncttrue
\mciteSetBstMidEndSepPunct{\mcitedefaultmidpunct}
{\mcitedefaultendpunct}{\mcitedefaultseppunct}\relax
\EndOfBibitem
\bibitem{LHCb-PAPER-2021-014}
LHCb collaboration, R.~Aaij {\em et~al.},
  \ifthenelse{\boolean{articletitles}}{\emph{{Branching fraction measurements
  of the rare $B^0_s \to \phi \mu^+\mu^-$ and $B^0_s \to f_2^\prime(1525)
  \mu^+\mu^-$ decays}},
  }{}\href{https://doi.org/10.1103/PhysRevLett.127.151801}{Phys.\ Rev.\ Lett.\
  \textbf{127} (2021) 151801},
  \href{http://arxiv.org/abs/2105.14007}{{\normalfont\ttfamily
  arXiv:2105.14007}}\relax
\mciteBstWouldAddEndPuncttrue
\mciteSetBstMidEndSepPunct{\mcitedefaultmidpunct}
{\mcitedefaultendpunct}{\mcitedefaultseppunct}\relax
\EndOfBibitem
\bibitem{CDF:2011tds}
CDF collaboration, T.~Aaltonen {\em et~al.},
  \ifthenelse{\boolean{articletitles}}{\emph{{Measurements of the angular
  distributions in the decays $B \to K^{(*)} \mu^+ \mu^-$ at CDF}},
  }{}\href{https://doi.org/10.1103/PhysRevLett.108.081807}{Phys.\ Rev.\ Lett.\
  \textbf{108} (2012) 081807},
  \href{http://arxiv.org/abs/1108.0695}{{\normalfont\ttfamily
  arXiv:1108.0695}}\relax
\mciteBstWouldAddEndPuncttrue
\mciteSetBstMidEndSepPunct{\mcitedefaultmidpunct}
{\mcitedefaultendpunct}{\mcitedefaultseppunct}\relax
\EndOfBibitem
\bibitem{CMS:2015bcy}
CMS collaboration, V.~Khachatryan {\em et~al.},
  \ifthenelse{\boolean{articletitles}}{\emph{{Angular analysis of the decay
  $B^0 \to K^{*0} \mu^+ \mu^-$ from pp collisions at $\sqrt s = 8$ TeV}},
  }{}\href{https://doi.org/10.1016/j.physletb.2015.12.020}{Phys.\ Lett.\ B
  \textbf{753} (2016) 424},
  \href{http://arxiv.org/abs/1507.08126}{{\normalfont\ttfamily
  arXiv:1507.08126}}\relax
\mciteBstWouldAddEndPuncttrue
\mciteSetBstMidEndSepPunct{\mcitedefaultmidpunct}
{\mcitedefaultendpunct}{\mcitedefaultseppunct}\relax
\EndOfBibitem
\bibitem{CMS:2017rzx}
CMS collaboration, A.~M. Sirunyan {\em et~al.},
  \ifthenelse{\boolean{articletitles}}{\emph{{Measurement of angular parameters
  from the decay $\mathrm{B}^0 \to \mathrm{K}^{*0} \mu^+ \mu^-$ in
  proton-proton collisions at $\sqrt{s} = $ 8 TeV}},
  }{}\href{https://doi.org/10.1016/j.physletb.2018.04.030}{Phys.\ Lett.\ B
  \textbf{781} (2018) 517},
  \href{http://arxiv.org/abs/1710.02846}{{\normalfont\ttfamily
  arXiv:1710.02846}}\relax
\mciteBstWouldAddEndPuncttrue
\mciteSetBstMidEndSepPunct{\mcitedefaultmidpunct}
{\mcitedefaultendpunct}{\mcitedefaultseppunct}\relax
\EndOfBibitem
\bibitem{CMS:2018qih}
CMS collaboration, A.~M. Sirunyan {\em et~al.},
  \ifthenelse{\boolean{articletitles}}{\emph{{Angular analysis of the decay
  B$^+\to$ K$^+\mu^+\mu^-$ in proton-proton collisions at $\sqrt{s} =$ 8 TeV}},
  }{}\href{https://doi.org/10.1103/PhysRevD.98.112011}{Phys.\ Rev.\ D
  \textbf{98} (2018) 112011},
  \href{http://arxiv.org/abs/1806.00636}{{\normalfont\ttfamily
  arXiv:1806.00636}}\relax
\mciteBstWouldAddEndPuncttrue
\mciteSetBstMidEndSepPunct{\mcitedefaultmidpunct}
{\mcitedefaultendpunct}{\mcitedefaultseppunct}\relax
\EndOfBibitem
\bibitem{ATLAS:2018gqc}
ATLAS collaboration, M.~Aaboud {\em et~al.},
  \ifthenelse{\boolean{articletitles}}{\emph{{Angular analysis of $B^0_d
  \rightarrow K^{*}\mu^+\mu^-$ decays in $pp$ collisions at $\sqrt{s}= 8$ TeV
  with the ATLAS detector}},
  }{}\href{https://doi.org/10.1007/JHEP10(2018)047}{JHEP \textbf{10} (2018)
  047}, \href{http://arxiv.org/abs/1805.04000}{{\normalfont\ttfamily
  arXiv:1805.04000}}\relax
\mciteBstWouldAddEndPuncttrue
\mciteSetBstMidEndSepPunct{\mcitedefaultmidpunct}
{\mcitedefaultendpunct}{\mcitedefaultseppunct}\relax
\EndOfBibitem
\bibitem{CMS:2020oqb}
CMS collaboration, A.~M. Sirunyan {\em et~al.},
  \ifthenelse{\boolean{articletitles}}{\emph{{Angular analysis of the decay
  B$^+$ $\to$ K$^*$(892)$^+\mu^+\mu^-$ in proton-proton collisions at $\sqrt{s}
  =$ 8 TeV}}, }{}\href{https://doi.org/10.1007/JHEP04(2021)124}{JHEP
  \textbf{04} (2021) 124},
  \href{http://arxiv.org/abs/2010.13968}{{\normalfont\ttfamily
  arXiv:2010.13968}}\relax
\mciteBstWouldAddEndPuncttrue
\mciteSetBstMidEndSepPunct{\mcitedefaultmidpunct}
{\mcitedefaultendpunct}{\mcitedefaultseppunct}\relax
\EndOfBibitem
\bibitem{LHCb-PAPER-2015-023}
LHCb collaboration, R.~Aaij {\em et~al.},
  \ifthenelse{\boolean{articletitles}}{\emph{{Angular analysis and differential
  branching fraction of the decay \mbox{\decay{\Bs}{\phi\mumu}}}},
  }{}\href{https://doi.org/10.1007/JHEP09(2015)179}{JHEP \textbf{09} (2015)
  179}, \href{http://arxiv.org/abs/1506.08777}{{\normalfont\ttfamily
  arXiv:1506.08777}}\relax
\mciteBstWouldAddEndPuncttrue
\mciteSetBstMidEndSepPunct{\mcitedefaultmidpunct}
{\mcitedefaultendpunct}{\mcitedefaultseppunct}\relax
\EndOfBibitem
\bibitem{LHCb-PAPER-2015-051}
LHCb collaboration, R.~Aaij {\em et~al.},
  \ifthenelse{\boolean{articletitles}}{\emph{{Angular analysis of the
  \mbox{\decay{\Bz}{\Kstarz\mumu}} decay using $3\invfb$ of integrated
  luminosity}}, }{}\href{https://doi.org/10.1007/JHEP02(2016)104}{JHEP
  \textbf{02} (2016) 104},
  \href{http://arxiv.org/abs/1512.04442}{{\normalfont\ttfamily
  arXiv:1512.04442}}\relax
\mciteBstWouldAddEndPuncttrue
\mciteSetBstMidEndSepPunct{\mcitedefaultmidpunct}
{\mcitedefaultendpunct}{\mcitedefaultseppunct}\relax
\EndOfBibitem
\bibitem{Belle:2016fev}
Belle collaboration, S.~Wehle {\em et~al.},
  \ifthenelse{\boolean{articletitles}}{\emph{{Lepton-flavor-dependent angular
  analysis of $B\to K^\ast \ell^+\ell^-$}},
  }{}\href{https://doi.org/10.1103/PhysRevLett.118.111801}{Phys.\ Rev.\ Lett.\
  \textbf{118} (2017) 111801},
  \href{http://arxiv.org/abs/1612.05014}{{\normalfont\ttfamily
  arXiv:1612.05014}}\relax
\mciteBstWouldAddEndPuncttrue
\mciteSetBstMidEndSepPunct{\mcitedefaultmidpunct}
{\mcitedefaultendpunct}{\mcitedefaultseppunct}\relax
\EndOfBibitem
\bibitem{LHCb-PAPER-2020-041}
LHCb collaboration, R.~Aaij {\em et~al.},
  \ifthenelse{\boolean{articletitles}}{\emph{{Angular analysis of the $B^{+}\to
  K^{\ast+}\mu^+\mu^-$ decay}},
  }{}\href{https://doi.org/10.1103/PhysRevLett.126.161802}{Phys.\ Rev.\ Lett.\
  \textbf{126} (2021) 161802},
  \href{http://arxiv.org/abs/2012.13241}{{\normalfont\ttfamily
  arXiv:2012.13241}}\relax
\mciteBstWouldAddEndPuncttrue
\mciteSetBstMidEndSepPunct{\mcitedefaultmidpunct}
{\mcitedefaultendpunct}{\mcitedefaultseppunct}\relax
\EndOfBibitem
\bibitem{LHCb-PAPER-2021-022}
LHCb collaboration, R.~Aaij {\em et~al.},
  \ifthenelse{\boolean{articletitles}}{\emph{{Angular analysis of the rare
  decay $\Bs \to \Pphi \mup \mun$}},
  }{}\href{https://doi.org/10.1007/JHEP11(2021)043}{JHEP \textbf{11} (2021)
  043}, \href{http://arxiv.org/abs/2107.13428}{{\normalfont\ttfamily
  arXiv:2107.13428}}\relax
\mciteBstWouldAddEndPuncttrue
\mciteSetBstMidEndSepPunct{\mcitedefaultmidpunct}
{\mcitedefaultendpunct}{\mcitedefaultseppunct}\relax
\EndOfBibitem
\bibitem{BaBar:2012mrf}
BaBar collaboration, J.~P. Lees {\em et~al.},
  \ifthenelse{\boolean{articletitles}}{\emph{{Measurement of branching
  fractions and rate asymmetries in the rare decays $B \to K^{(*)} l^+ l^-$}},
  }{}\href{https://doi.org/10.1103/PhysRevD.86.032012}{Phys.\ Rev.\
  \textbf{D86} (2012) 032012},
  \href{http://arxiv.org/abs/1204.3933}{{\normalfont\ttfamily
  arXiv:1204.3933}}\relax
\mciteBstWouldAddEndPuncttrue
\mciteSetBstMidEndSepPunct{\mcitedefaultmidpunct}
{\mcitedefaultendpunct}{\mcitedefaultseppunct}\relax
\EndOfBibitem
\bibitem{LHCb-PAPER-2019-009}
LHCb collaboration, R.~Aaij {\em et~al.},
  \ifthenelse{\boolean{articletitles}}{\emph{{Search for lepton-universality
  violation in \mbox{\decay{\Bp}{\Kp\ellell}} decays}},
  }{}\href{https://doi.org/10.1103/PhysRevLett.122.191801}{Phys.\ Rev.\ Lett.\
  \textbf{122} (2019) 191801},
  \href{http://arxiv.org/abs/1903.09252}{{\normalfont\ttfamily
  arXiv:1903.09252}}\relax
\mciteBstWouldAddEndPuncttrue
\mciteSetBstMidEndSepPunct{\mcitedefaultmidpunct}
{\mcitedefaultendpunct}{\mcitedefaultseppunct}\relax
\EndOfBibitem
\bibitem{BELLE:2019xld}
Belle collaboration, S.~Choudhury {\em et~al.},
  \ifthenelse{\boolean{articletitles}}{\emph{{Test of lepton flavor
  universality and search for lepton flavor violation in $B \rightarrow K\ell
  \ell$ decays}}, }{}\href{https://doi.org/10.1007/JHEP03(2021)105}{JHEP
  \textbf{03} (2021) 105},
  \href{http://arxiv.org/abs/1908.01848}{{\normalfont\ttfamily
  arXiv:1908.01848}}\relax
\mciteBstWouldAddEndPuncttrue
\mciteSetBstMidEndSepPunct{\mcitedefaultmidpunct}
{\mcitedefaultendpunct}{\mcitedefaultseppunct}\relax
\EndOfBibitem
\bibitem{Belle:2019oag}
Belle collaboration, A.~Abdesselam {\em et~al.},
  \ifthenelse{\boolean{articletitles}}{\emph{{Test of lepton-flavor
  universality in ${B\to K^\ast\ell^+\ell^-}$ decays at Belle}},
  }{}\href{https://doi.org/10.1103/PhysRevLett.126.161801}{Phys.\ Rev.\ Lett.\
  \textbf{126} (2021) 161801},
  \href{http://arxiv.org/abs/1904.02440}{{\normalfont\ttfamily
  arXiv:1904.02440}}\relax
\mciteBstWouldAddEndPuncttrue
\mciteSetBstMidEndSepPunct{\mcitedefaultmidpunct}
{\mcitedefaultendpunct}{\mcitedefaultseppunct}\relax
\EndOfBibitem
\bibitem{LHCb-PAPER-2021-038}
LHCb collaboration, R.~Aaij {\em et~al.},
  \ifthenelse{\boolean{articletitles}}{\emph{{Tests of lepton universality
  using \mbox{$B^0 \to \KS \ell^+\ell^-$} and \mbox{$B^+\to K^{*+}\ell^+
  \ell^-$} decays}},
  }{}\href{https://doi.org/10.1103/PhysRevLett.128.191802}{Phys.\ Rev.\ Lett.\
  \textbf{128} (2022) 191802},
  \href{http://arxiv.org/abs/2110.09501}{{\normalfont\ttfamily
  arXiv:2110.09501}}\relax
\mciteBstWouldAddEndPuncttrue
\mciteSetBstMidEndSepPunct{\mcitedefaultmidpunct}
{\mcitedefaultendpunct}{\mcitedefaultseppunct}\relax
\EndOfBibitem
\bibitem{LHCb:2022qnv}
LHCb collaboration, \ifthenelse{\boolean{articletitles}}{\emph{{Test of lepton
  universality in $b\rightarrow s\ell^+ \ell^-$ decays}},
  }{}\href{http://arxiv.org/abs/2212.09152}{{\normalfont\ttfamily
  arXiv:2212.09152}}\relax
\mciteBstWouldAddEndPuncttrue
\mciteSetBstMidEndSepPunct{\mcitedefaultmidpunct}
{\mcitedefaultendpunct}{\mcitedefaultseppunct}\relax
\EndOfBibitem
\bibitem{LHCb:2022zom}
LHCb collaboration, \ifthenelse{\boolean{articletitles}}{\emph{{Measurement of
  lepton universality parameters in $B^+\to K^+\ell^+\ell^-$ and $B^0\to
  K^{*0}\ell^+\ell^-$ decays}},
  }{}\href{http://arxiv.org/abs/2212.09153}{{\normalfont\ttfamily
  arXiv:2212.09153}}\relax
\mciteBstWouldAddEndPuncttrue
\mciteSetBstMidEndSepPunct{\mcitedefaultmidpunct}
{\mcitedefaultendpunct}{\mcitedefaultseppunct}\relax
\EndOfBibitem
\bibitem{Feldmann:2021zta}
T.~Feldmann, \ifthenelse{\boolean{articletitles}}{\emph{{Theory: Angular
  Distributions in Rare B Decays}},
  }{}\href{https://doi.org/10.22323/1.391.0018}{PoS \textbf{BEAUTY2020} (2021)
  018}, \href{http://arxiv.org/abs/2101.04314}{{\normalfont\ttfamily
  arXiv:2101.04314}}\relax
\mciteBstWouldAddEndPuncttrue
\mciteSetBstMidEndSepPunct{\mcitedefaultmidpunct}
{\mcitedefaultendpunct}{\mcitedefaultseppunct}\relax
\EndOfBibitem
\bibitem{LHCb-PAPER-2015-009}
LHCb collaboration, R.~Aaij {\em et~al.},
  \ifthenelse{\boolean{articletitles}}{\emph{{Differential branching fraction
  and angular analysis of \mbox{\decay{\Lb}{\Lz\mumu}} decays}},
  }{}\href{https://doi.org/10.1007/JHEP06(2015)115}{JHEP \textbf{06} (2015)
  115}, Erratum \href{https://doi.org/10.1007/JHEP09(2018)145}{ibid.\
  \textbf{09} (2018) 145},
  \href{http://arxiv.org/abs/1503.07138}{{\normalfont\ttfamily
  arXiv:1503.07138}}\relax
\mciteBstWouldAddEndPuncttrue
\mciteSetBstMidEndSepPunct{\mcitedefaultmidpunct}
{\mcitedefaultendpunct}{\mcitedefaultseppunct}\relax
\EndOfBibitem
\bibitem{Detmold:2016pkz}
W.~Detmold and S.~Meinel, \ifthenelse{\boolean{articletitles}}{\emph{{$\Lb \to
  \Lz \ell^+ \ell^-$ form factors, differential branching fraction, and angular
  observables from lattice QCD with relativistic $b$ quarks}},
  }{}\href{https://doi.org/10.1103/PhysRevD.93.074501}{Phys.\ Rev.\
  \textbf{D93} (2016) 074501},
  \href{http://arxiv.org/abs/1602.01399}{{\normalfont\ttfamily
  arXiv:1602.01399}}\relax
\mciteBstWouldAddEndPuncttrue
\mciteSetBstMidEndSepPunct{\mcitedefaultmidpunct}
{\mcitedefaultendpunct}{\mcitedefaultseppunct}\relax
\EndOfBibitem
\bibitem{Blake:2019guk}
T.~Blake, S.~Meinel, and D.~van Dyk,
  \ifthenelse{\boolean{articletitles}}{\emph{{Bayesian analysis of $b\to
  s\mu^+\mu^-$ Wilson coefficients using the full angular distribution of
  $\Lb\to \PLambda(\to p\, \pi^-)\mu^+\mu^-$ decays}},
  }{}\href{https://doi.org/10.1103/PhysRevD.101.035023}{Phys.\ Rev.\
  \textbf{D101} (2020) 035023},
  \href{http://arxiv.org/abs/1912.05811}{{\normalfont\ttfamily
  arXiv:1912.05811}}\relax
\mciteBstWouldAddEndPuncttrue
\mciteSetBstMidEndSepPunct{\mcitedefaultmidpunct}
{\mcitedefaultendpunct}{\mcitedefaultseppunct}\relax
\EndOfBibitem
\bibitem{LHCb-PAPER-2016-059}
LHCb collaboration, R.~Aaij {\em et~al.},
  \ifthenelse{\boolean{articletitles}}{\emph{{Observation of the decay
  \mbox{\decay{\Lb}{\proton\Km\mumu}} and search for \CP violation}},
  }{}\href{https://doi.org/10.1007/JHEP06(2017)108}{JHEP \textbf{06} (2017)
  108}, \href{http://arxiv.org/abs/1703.00256}{{\normalfont\ttfamily
  arXiv:1703.00256}}\relax
\mciteBstWouldAddEndPuncttrue
\mciteSetBstMidEndSepPunct{\mcitedefaultmidpunct}
{\mcitedefaultendpunct}{\mcitedefaultseppunct}\relax
\EndOfBibitem
\bibitem{LHCb-PAPER-2019-040}
LHCb collaboration, R.~Aaij {\em et~al.},
  \ifthenelse{\boolean{articletitles}}{\emph{{Test of lepton universality using
  \mbox{\decay{\Lb}{p\Km\ellell}} decays}},
  }{}\href{https://doi.org/10.1007/JHEP05(2020)040}{JHEP \textbf{05} (2020)
  040}, \href{http://arxiv.org/abs/1912.08139}{{\normalfont\ttfamily
  arXiv:1912.08139}}\relax
\mciteBstWouldAddEndPuncttrue
\mciteSetBstMidEndSepPunct{\mcitedefaultmidpunct}
{\mcitedefaultendpunct}{\mcitedefaultseppunct}\relax
\EndOfBibitem
\bibitem{Sarantsev:2019xxm}
A.~V. Sarantsev {\em et~al.},
  \ifthenelse{\boolean{articletitles}}{\emph{{Hyperon II: Properties of excited
  hyperons}}, }{}\href{https://doi.org/10.1140/epja/i2019-12880-5}{Eur.\ Phys.\
  J.\  \textbf{A55} (2019) 180},
  \href{http://arxiv.org/abs/1907.13387}{{\normalfont\ttfamily
  arXiv:1907.13387}}\relax
\mciteBstWouldAddEndPuncttrue
\mciteSetBstMidEndSepPunct{\mcitedefaultmidpunct}
{\mcitedefaultendpunct}{\mcitedefaultseppunct}\relax
\EndOfBibitem
\bibitem{Meinel:2020owd}
S.~Meinel and G.~Rendon, \ifthenelse{\boolean{articletitles}}{\emph{{$\Lb \to
  \Lz^*(1520)\ell^+\ell^-$ form factors from lattice QCD}},
  }{}\href{https://doi.org/10.1103/PhysRevD.103.074505}{Phys.\ Rev.\
  \textbf{D103} (2021) 074505},
  \href{http://arxiv.org/abs/2009.09313}{{\normalfont\ttfamily
  arXiv:2009.09313}}\relax
\mciteBstWouldAddEndPuncttrue
\mciteSetBstMidEndSepPunct{\mcitedefaultmidpunct}
{\mcitedefaultendpunct}{\mcitedefaultseppunct}\relax
\EndOfBibitem
\bibitem{PDG2022}
Particle Data Group, R.~L. Workman {\em et~al.},
  \ifthenelse{\boolean{articletitles}}{\emph{{\href{http://pdg.lbl.gov/}{Review
  of particle physics}}}, }{}\href{https://doi.org/10.1093/ptep/ptac097}{Prog.\
  Theor.\ Exp.\ Phys.\  \textbf{2022} (2022) 083C01}\relax
\mciteBstWouldAddEndPuncttrue
\mciteSetBstMidEndSepPunct{\mcitedefaultmidpunct}
{\mcitedefaultendpunct}{\mcitedefaultseppunct}\relax
\EndOfBibitem
\bibitem{LHCb-DP-2008-001}
LHCb collaboration, A.~A. Alves~Jr.\ {\em et~al.},
  \ifthenelse{\boolean{articletitles}}{\emph{{The \lhcb detector at the LHC}},
  }{}\href{https://doi.org/10.1088/1748-0221/3/08/S08005}{JINST \textbf{3}
  (2008) S08005}\relax
\mciteBstWouldAddEndPuncttrue
\mciteSetBstMidEndSepPunct{\mcitedefaultmidpunct}
{\mcitedefaultendpunct}{\mcitedefaultseppunct}\relax
\EndOfBibitem
\bibitem{LHCb-DP-2014-002}
LHCb collaboration, R.~Aaij {\em et~al.},
  \ifthenelse{\boolean{articletitles}}{\emph{{LHCb detector performance}},
  }{}\href{https://doi.org/10.1142/S0217751X15300227}{Int.\ J.\ Mod.\ Phys.\
  \textbf{A30} (2015) 1530022},
  \href{http://arxiv.org/abs/1412.6352}{{\normalfont\ttfamily
  arXiv:1412.6352}}\relax
\mciteBstWouldAddEndPuncttrue
\mciteSetBstMidEndSepPunct{\mcitedefaultmidpunct}
{\mcitedefaultendpunct}{\mcitedefaultseppunct}\relax
\EndOfBibitem
\bibitem{LHCb-DP-2012-004}
R.~Aaij {\em et~al.}, \ifthenelse{\boolean{articletitles}}{\emph{{The \lhcb
  trigger and its performance in 2011}},
  }{}\href{https://doi.org/10.1088/1748-0221/8/04/P04022}{JINST \textbf{8}
  (2013) P04022}, \href{http://arxiv.org/abs/1211.3055}{{\normalfont\ttfamily
  arXiv:1211.3055}}\relax
\mciteBstWouldAddEndPuncttrue
\mciteSetBstMidEndSepPunct{\mcitedefaultmidpunct}
{\mcitedefaultendpunct}{\mcitedefaultseppunct}\relax
\EndOfBibitem
\bibitem{LHCb-DP-2019-001}
R.~Aaij {\em et~al.}, \ifthenelse{\boolean{articletitles}}{\emph{{Design and
  performance of the LHCb trigger and full real-time reconstruction in Run 2 of
  the LHC}}, }{}\href{https://doi.org/10.1088/1748-0221/14/04/P04013}{JINST
  \textbf{14} (2019) P04013},
  \href{http://arxiv.org/abs/1812.10790}{{\normalfont\ttfamily
  arXiv:1812.10790}}\relax
\mciteBstWouldAddEndPuncttrue
\mciteSetBstMidEndSepPunct{\mcitedefaultmidpunct}
{\mcitedefaultendpunct}{\mcitedefaultseppunct}\relax
\EndOfBibitem
\bibitem{LHCb-DP-2013-001}
F.~Archilli {\em et~al.},
  \ifthenelse{\boolean{articletitles}}{\emph{{Performance of the muon
  identification at LHCb}},
  }{}\href{https://doi.org/10.1088/1748-0221/8/10/P10020}{JINST \textbf{8}
  (2013) P10020}, \href{http://arxiv.org/abs/1306.0249}{{\normalfont\ttfamily
  arXiv:1306.0249}}\relax
\mciteBstWouldAddEndPuncttrue
\mciteSetBstMidEndSepPunct{\mcitedefaultmidpunct}
{\mcitedefaultendpunct}{\mcitedefaultseppunct}\relax
\EndOfBibitem
\bibitem{Sjostrand:2007gs}
T.~Sj\"{o}strand, S.~Mrenna, and P.~Skands,
  \ifthenelse{\boolean{articletitles}}{\emph{{A brief introduction to PYTHIA
  8.1}}, }{}\href{https://doi.org/10.1016/j.cpc.2008.01.036}{Comput.\ Phys.\
  Commun.\  \textbf{178} (2008) 852},
  \href{http://arxiv.org/abs/0710.3820}{{\normalfont\ttfamily
  arXiv:0710.3820}}\relax
\mciteBstWouldAddEndPuncttrue
\mciteSetBstMidEndSepPunct{\mcitedefaultmidpunct}
{\mcitedefaultendpunct}{\mcitedefaultseppunct}\relax
\EndOfBibitem
\bibitem{Sjostrand:2006za}
T.~Sj\"{o}strand, S.~Mrenna, and P.~Skands,
  \ifthenelse{\boolean{articletitles}}{\emph{{PYTHIA 6.4 physics and manual}},
  }{}\href{https://doi.org/10.1088/1126-6708/2006/05/026}{JHEP \textbf{05}
  (2006) 026}, \href{http://arxiv.org/abs/hep-ph/0603175}{{\normalfont\ttfamily
  arXiv:hep-ph/0603175}}\relax
\mciteBstWouldAddEndPuncttrue
\mciteSetBstMidEndSepPunct{\mcitedefaultmidpunct}
{\mcitedefaultendpunct}{\mcitedefaultseppunct}\relax
\EndOfBibitem
\bibitem{LHCb-PROC-2010-056}
I.~Belyaev {\em et~al.}, \ifthenelse{\boolean{articletitles}}{\emph{{Handling
  of the generation of primary events in Gauss, the LHCb simulation
  framework}}, }{}\href{https://doi.org/10.1088/1742-6596/331/3/032047}{J.\
  Phys.\ Conf.\ Ser.\  \textbf{331} (2011) 032047}\relax
\mciteBstWouldAddEndPuncttrue
\mciteSetBstMidEndSepPunct{\mcitedefaultmidpunct}
{\mcitedefaultendpunct}{\mcitedefaultseppunct}\relax
\EndOfBibitem
\bibitem{Lange:2001uf}
D.~J. Lange, \ifthenelse{\boolean{articletitles}}{\emph{{The EvtGen particle
  decay simulation package}},
  }{}\href{https://doi.org/10.1016/S0168-9002(01)00089-4}{Nucl.\ Instrum.\
  Meth.\  \textbf{A462} (2001) 152}\relax
\mciteBstWouldAddEndPuncttrue
\mciteSetBstMidEndSepPunct{\mcitedefaultmidpunct}
{\mcitedefaultendpunct}{\mcitedefaultseppunct}\relax
\EndOfBibitem
\bibitem{Golonka:2005pn}
P.~Golonka and Z.~Was, \ifthenelse{\boolean{articletitles}}{\emph{{PHOTOS Monte
  Carlo: A precision tool for QED corrections in $Z$ and $W$ decays}},
  }{}\href{https://doi.org/10.1140/epjc/s2005-02396-4}{Eur.\ Phys.\ J.\
  \textbf{C45} (2006) 97},
  \href{http://arxiv.org/abs/hep-ph/0506026}{{\normalfont\ttfamily
  arXiv:hep-ph/0506026}}\relax
\mciteBstWouldAddEndPuncttrue
\mciteSetBstMidEndSepPunct{\mcitedefaultmidpunct}
{\mcitedefaultendpunct}{\mcitedefaultseppunct}\relax
\EndOfBibitem
\bibitem{Allison:2006ve}
Geant4 collaboration, J.~Allison {\em et~al.},
  \ifthenelse{\boolean{articletitles}}{\emph{{Geant4 developments and
  applications}}, }{}\href{https://doi.org/10.1109/TNS.2006.869826}{IEEE
  Trans.\ Nucl.\ Sci.\  \textbf{53} (2006) 270}\relax
\mciteBstWouldAddEndPuncttrue
\mciteSetBstMidEndSepPunct{\mcitedefaultmidpunct}
{\mcitedefaultendpunct}{\mcitedefaultseppunct}\relax
\EndOfBibitem
\bibitem{Agostinelli:2002hh}
Geant4 collaboration, S.~Agostinelli {\em et~al.},
  \ifthenelse{\boolean{articletitles}}{\emph{{Geant4: A simulation toolkit}},
  }{}\href{https://doi.org/10.1016/S0168-9002(03)01368-8}{Nucl.\ Instrum.\
  Meth.\  \textbf{A506} (2003) 250}\relax
\mciteBstWouldAddEndPuncttrue
\mciteSetBstMidEndSepPunct{\mcitedefaultmidpunct}
{\mcitedefaultendpunct}{\mcitedefaultseppunct}\relax
\EndOfBibitem
\bibitem{LHCb-PROC-2011-006}
M.~Clemencic {\em et~al.}, \ifthenelse{\boolean{articletitles}}{\emph{{The
  \lhcb simulation application, Gauss: Design, evolution and experience}},
  }{}\href{https://doi.org/10.1088/1742-6596/331/3/032023}{J.\ Phys.\ Conf.\
  Ser.\  \textbf{331} (2011) 032023}\relax
\mciteBstWouldAddEndPuncttrue
\mciteSetBstMidEndSepPunct{\mcitedefaultmidpunct}
{\mcitedefaultendpunct}{\mcitedefaultseppunct}\relax
\EndOfBibitem
\bibitem{LHCb-PAPER-2015-029}
LHCb collaboration, R.~Aaij {\em et~al.},
  \ifthenelse{\boolean{articletitles}}{\emph{{Observation of $\jpsi\proton$
  resonances consistent with pentaquark states in
  \mbox{\decay{\Lb}{\jpsi\proton\Km}} decays}},
  }{}\href{https://doi.org/10.1103/PhysRevLett.115.072001}{Phys.\ Rev.\ Lett.\
  \textbf{115} (2015) 072001},
  \href{http://arxiv.org/abs/1507.03414}{{\normalfont\ttfamily
  arXiv:1507.03414}}\relax
\mciteBstWouldAddEndPuncttrue
\mciteSetBstMidEndSepPunct{\mcitedefaultmidpunct}
{\mcitedefaultendpunct}{\mcitedefaultseppunct}\relax
\EndOfBibitem
\bibitem{Breiman}
L.~Breiman, J.~H. Friedman, R.~A. Olshen, and C.~J. Stone, {\em Classification
  and regression trees}, Wadsworth international group, Belmont, California,
  USA, 1984\relax
\mciteBstWouldAddEndPuncttrue
\mciteSetBstMidEndSepPunct{\mcitedefaultmidpunct}
{\mcitedefaultendpunct}{\mcitedefaultseppunct}\relax
\EndOfBibitem
\bibitem{Roe:2004na}
B.~P. Roe {\em et~al.}, \ifthenelse{\boolean{articletitles}}{\emph{{Boosted
  decision trees, an alternative to artificial neural networks}},
  }{}\href{https://doi.org/10.1016/j.nima.2004.12.018}{Nucl.\ Instrum.\ Meth.\
  \textbf{A543} (2005) 577},
  \href{http://arxiv.org/abs/physics/0408124}{{\normalfont\ttfamily
  arXiv:physics/0408124}}\relax
\mciteBstWouldAddEndPuncttrue
\mciteSetBstMidEndSepPunct{\mcitedefaultmidpunct}
{\mcitedefaultendpunct}{\mcitedefaultseppunct}\relax
\EndOfBibitem
\bibitem{Hocker:2007ht}
H.~Voss, A.~Hoecker, J.~Stelzer, and F.~Tegenfeldt,
  \ifthenelse{\boolean{articletitles}}{\emph{{TMVA - Toolkit for Multivariate
  Data Analysis with ROOT}}, }{}\href{https://doi.org/10.22323/1.050.0040}{PoS
  \textbf{ACAT} (2007) 040}\relax
\mciteBstWouldAddEndPuncttrue
\mciteSetBstMidEndSepPunct{\mcitedefaultmidpunct}
{\mcitedefaultendpunct}{\mcitedefaultseppunct}\relax
\EndOfBibitem
\bibitem{LHCb-DP-2018-001}
R.~Aaij {\em et~al.}, \ifthenelse{\boolean{articletitles}}{\emph{{Selection and
  processing of calibration samples to measure the particle identification
  performance of the LHCb experiment in Run 2}},
  }{}\href{https://doi.org/10.1140/epjti/s40485-019-0050-z}{Eur.\ Phys.\ J.\
  Tech.\ Instr.\  \textbf{6} (2019) 1},
  \href{http://arxiv.org/abs/1803.00824}{{\normalfont\ttfamily
  arXiv:1803.00824}}\relax
\mciteBstWouldAddEndPuncttrue
\mciteSetBstMidEndSepPunct{\mcitedefaultmidpunct}
{\mcitedefaultendpunct}{\mcitedefaultseppunct}\relax
\EndOfBibitem
\bibitem{Santos:2013gra}
D.~Mart{\'\i}nez~Santos and F.~Dupertuis,
  \ifthenelse{\boolean{articletitles}}{\emph{{Mass distributions marginalized
  over per-event errors}},
  }{}\href{https://doi.org/10.1016/j.nima.2014.06.081}{Nucl.\ Instrum.\ Meth.\
  \textbf{A764} (2014) 150},
  \href{http://arxiv.org/abs/1312.5000}{{\normalfont\ttfamily
  arXiv:1312.5000}}\relax
\mciteBstWouldAddEndPuncttrue
\mciteSetBstMidEndSepPunct{\mcitedefaultmidpunct}
{\mcitedefaultendpunct}{\mcitedefaultseppunct}\relax
\EndOfBibitem
\bibitem{sup}
See supplemental material at [link inserted by publisher] for the additional
  figures\relax
\mciteBstWouldAddEndPuncttrue
\mciteSetBstMidEndSepPunct{\mcitedefaultmidpunct}
{\mcitedefaultendpunct}{\mcitedefaultseppunct}\relax
\EndOfBibitem
\bibitem{Pivk:2004ty}
M.~Pivk and F.~R. Le~Diberder,
  \ifthenelse{\boolean{articletitles}}{\emph{{sPlot: A statistical tool to
  unfold data distributions}},
  }{}\href{https://doi.org/10.1016/j.nima.2005.08.106}{Nucl.\ Instrum.\ Meth.\
  \textbf{A555} (2005) 356},
  \href{http://arxiv.org/abs/physics/0402083}{{\normalfont\ttfamily
  arXiv:physics/0402083}}\relax
\mciteBstWouldAddEndPuncttrue
\mciteSetBstMidEndSepPunct{\mcitedefaultmidpunct}
{\mcitedefaultendpunct}{\mcitedefaultseppunct}\relax
\EndOfBibitem
\bibitem{Xie:2009rka}
Y.~Xie, \ifthenelse{\boolean{articletitles}}{\emph{{sFit: a method for
  background subtraction in maximum likelihood fit}},
  }{}\href{http://arxiv.org/abs/0905.0724}{{\normalfont\ttfamily
  arXiv:0905.0724}}\relax
\mciteBstWouldAddEndPuncttrue
\mciteSetBstMidEndSepPunct{\mcitedefaultmidpunct}
{\mcitedefaultendpunct}{\mcitedefaultseppunct}\relax
\EndOfBibitem
\bibitem{Langenbruch:2019nwe}
C.~Langenbruch, \ifthenelse{\boolean{articletitles}}{\emph{{Parameter
  uncertainties in weighted unbinned maximum likelihood fits}},
  }{}\href{https://doi.org/10.1140/epjc/s10052-022-10254-8}{Eur.\ Phys.\ J.\
  \textbf{C82} (2022) 393},
  \href{http://arxiv.org/abs/1911.01303}{{\normalfont\ttfamily
  arXiv:1911.01303}}\relax
\mciteBstWouldAddEndPuncttrue
\mciteSetBstMidEndSepPunct{\mcitedefaultmidpunct}
{\mcitedefaultendpunct}{\mcitedefaultseppunct}\relax
\EndOfBibitem
\bibitem{Hassanvand:2012dn}
M.~Hassanvand, S.~Z. Kalantari, Y.~Akaishi, and T.~Yamazaki,
  \ifthenelse{\boolean{articletitles}}{\emph{{Theoretical analysis of
  $\Lambda(1405)\to(\Sigma \pi)^0$ mass spectra produced in $p+p \to
  p+\Lambda(1405)+K^+$ reactions}},
  }{}\href{https://doi.org/10.1103/PhysRevC.87.055202}{Phys.\ Rev.\
  \textbf{C87} (2013) 055202}, Addendum
  \href{https://doi.org/10.1103/PhysRevC.88.019905}{ibid.\   \textbf{C88}
  (2013) 019905}, \href{http://arxiv.org/abs/1210.7725}{{\normalfont\ttfamily
  arXiv:1210.7725}}\relax
\mciteBstWouldAddEndPuncttrue
\mciteSetBstMidEndSepPunct{\mcitedefaultmidpunct}
{\mcitedefaultendpunct}{\mcitedefaultseppunct}\relax
\EndOfBibitem
\bibitem{Blatt:1952ije}
J.~M. Blatt and V.~F. Weisskopf, {\em {Theoretical nuclear physics}},
  \href{https://doi.org/10.1007/978-1-4612-9959-2}{ Springer, New York,
  1952}\relax
\mciteBstWouldAddEndPuncttrue
\mciteSetBstMidEndSepPunct{\mcitedefaultmidpunct}
{\mcitedefaultendpunct}{\mcitedefaultseppunct}\relax
\EndOfBibitem
\bibitem{Skwarnicki:1986xj}
T.~Skwarnicki, {\em {A study of the radiative cascade transitions between the
  Upsilon-prime and Upsilon resonances}}, PhD thesis, Institute of Nuclear
  Physics, Krakow, 1986,
  {\href{http://inspirehep.net/record/230779/}{DESY-F31-86-02}}\relax
\mciteBstWouldAddEndPuncttrue
\mciteSetBstMidEndSepPunct{\mcitedefaultmidpunct}
{\mcitedefaultendpunct}{\mcitedefaultseppunct}\relax
\EndOfBibitem
\bibitem{smith2010chebyshev}
J.~Smith, \ifthenelse{\boolean{articletitles}}{\emph{The chebyshev function:
  Theory and applications}, }{}Mathematical Review \textbf{42} (2010) 135\relax
\mciteBstWouldAddEndPuncttrue
\mciteSetBstMidEndSepPunct{\mcitedefaultmidpunct}
{\mcitedefaultendpunct}{\mcitedefaultseppunct}\relax
\EndOfBibitem
\bibitem{Descotes-Genon:2019dbw}
S.~Descotes-Genon and M.~Novoa-Brunet,
  \ifthenelse{\boolean{articletitles}}{\emph{{Angular analysis of the rare
  decay $\Lambda_b\to \Lambda(1520)(\to N\bar{K})\ell^+\ell^-$}},
  }{}\href{https://doi.org/10.1007/JHEP06(2019)136}{JHEP \textbf{06} (2019)
  136}, Erratum \href{https://doi.org/10.1007/JHEP06(2020)102}{ibid.\
  \textbf{06} (2020) 102},
  \href{http://arxiv.org/abs/1903.00448}{{\normalfont\ttfamily
  arXiv:1903.00448}}\relax
\mciteBstWouldAddEndPuncttrue
\mciteSetBstMidEndSepPunct{\mcitedefaultmidpunct}
{\mcitedefaultendpunct}{\mcitedefaultseppunct}\relax
\EndOfBibitem
\bibitem{Li:2022nim}
Y.-S. Li, S.-P. Jin, J.~Gao, and X.~Liu,
  \ifthenelse{\boolean{articletitles}}{\emph{{The angular analysis of
  $\Lambda_b\to\Lambda(1520)(\to N\bar{K})\ell^+\ell^-$ decay}},
  }{}\href{http://arxiv.org/abs/2210.04640}{{\normalfont\ttfamily
  arXiv:2210.04640}}\relax
\mciteBstWouldAddEndPuncttrue
\mciteSetBstMidEndSepPunct{\mcitedefaultmidpunct}
{\mcitedefaultendpunct}{\mcitedefaultseppunct}\relax
\EndOfBibitem
\bibitem{Amhis:2022vcd}
Y.~Amhis, M.~Bordone, and M.~Reboud,
  \ifthenelse{\boolean{articletitles}}{\emph{{Dispersive analysis of $\Lambda_b
  \to \Lambda(1520)$ local form factors}},
  }{}\href{http://arxiv.org/abs/2208.08937}{{\normalfont\ttfamily
  arXiv:2208.08937}}\relax
\mciteBstWouldAddEndPuncttrue
\mciteSetBstMidEndSepPunct{\mcitedefaultmidpunct}
{\mcitedefaultendpunct}{\mcitedefaultseppunct}\relax
\EndOfBibitem
\bibitem{Meinel:2021mdj}
S.~Meinel and G.~Rendon,
  \ifthenelse{\boolean{articletitles}}{\emph{{$\PLambda_{c}\to\PLambda^{*}(1520)$
  form factors from lattice QCD and improved analysis of the
  $\PLambda_{b}\to\PLambda^{*}(1520)$ and
  $\PLambda_{b}\to\PLambda^{*}_{c}(2595,2625)$ form factors}},
  }{}\href{https://doi.org/10.1103/PhysRevD.105.054511}{Phys.\ Rev.\
  \textbf{D105} (2022) 054511},
  \href{http://arxiv.org/abs/2107.13140}{{\normalfont\ttfamily
  arXiv:2107.13140}}\relax
\mciteBstWouldAddEndPuncttrue
\mciteSetBstMidEndSepPunct{\mcitedefaultmidpunct}
{\mcitedefaultendpunct}{\mcitedefaultseppunct}\relax
\EndOfBibitem
\end{mcitethebibliography}
\end{document}